\documentclass[preprint,prc,showpacs,preprintnumbers,amsmath,amssymb,floatfix]{revtex4}
\usepackage{amsmath}

\usepackage{graphicx}
\usepackage{dcolumn}
\usepackage{bm}
\usepackage{ulem} 
\usepackage[usenames]{color}
\usepackage{epstopdf}
\usepackage{epsfig}
\usepackage{float}
\usepackage{subfigure}
\usepackage{multirow}
\usepackage[version=3]{mhchem}

\usepackage{hyperref}
\usepackage{array}
\usepackage[titletoc]{appendix}

\newcommand{\nc}{\newcommand}       
\nc{\vc}[1] {\mbox{\boldmath $#1$}} 
\nc{\del}       {\partial}              
\nc{\bra}       {\langle}               
\nc{\ket}       {\rangle}               
\nc{\bras}[1]   {\langle #1|}           
\nc{\kets}[1]   {|#1\rangle}            
\nc{\mapleft}[1]{           
 \smash{\mathop{\,          %
  \hbox to 1.5cm{\rightarrowfill}\, }\limits_{#1}}}
\nc{\beq}     {\begin{eqnarray}} \nc{\eeq}    {\end{eqnarray}}
\nc{\nn}      {\\\nonumber} \nc{\vs}      {\vspace{-0.275cm}}
\nc{\fra}    {\frac{1}{2}}
\nc{\mb}        {\mathbf}

\begin{document}

\preprint{}

\title{The charge-dependent Bonn potentials with pseudovector pion-nucleon coupling}

\author{Chencan Wang}
\affiliation{School of Physics, Nankai University, Tianjin 300071,  China}
\author{Jinniu Hu}
\email{hujinniu@nankai.edu.cn}
\affiliation{School of Physics, Nankai University, Tianjin 300071,  China}
\author{Ying Zhang}
\affiliation{Department of Physics, Faculty of Science, Tianjin University, Tianjin 300072, China}
\author{Hong Shen}
\affiliation{School of Physics, Nankai University, Tianjin 300071,  China}

\date{\today}
\begin{abstract}
To apply the high-precision realistic nucleon-nucleon ($NN$) potentials on  the investigations of relativistic many-body methods, the new versions of	charge-dependent Bonn (CD-Bonn) $NN$ potential are constructed within the pseudovector pion-nucleon coupling instead of the  pseudoscalar type in the original CD-Bonn potential worked out by Machleidt [Phys. Rev. C 63, 024001 (2001)]. Two effective scalar mesons are introduced, whose coupling constants with nucleon are independently determined at each partial wave for total angular momentum $J\leq 4$, to describe the charge dependence of $NN$ scattering data precisely, while the coupling constants between vector, pseudovector mesons and nucleon are  identical in all channels. Three revised CD-Bonn potentials adopting the pseudovector pion-nucleon couplings (pvCD-Bonn) are generated by fitting the Nijmegen PWA phase shift data and deuteron binding energy with different pion-nucleon coupling strengths, which can reproduce the phase shifts at spin-single channels and low-energy $NN$ scattering parameters very well, and provide the significantly different mixing parameters at spin-triplet channels. Furthermore, the $D$-state probabilities of deuteron from these potentials range from $4.22\%$ to  $6.05\%$. It demonstrates that these potentials contain different components of tensor force, which will be useful to discuss the roles of tensor force in nuclear few-body and many-body systems.  
\end{abstract}



\maketitle

\section{Introduction}
The nucleon-nucleon ($NN$) interaction is the most essential physical quantity to interpret the world of nuclear physics, which determines how the protons and neutrons compose a complex quantum system. It is regarded as a residual effect of strong interaction at low energy region, where the quantum chromodynamics (QCD) theory cannot be solved perturbatively. Therefore, the $NN$ interaction is usually described by the meson-exchange model proposed by Yukawa firstly~\cite{Yukawa35}. In the 1960s, heavier bosons were included in this scheme to deal with the intermediate- and short-range regions of $NN$ interaction.  Generally speaking, there are two kinds of nuclear interactions. The first kind is constructed based on the free $NN$ scattering data, such as differential cross section and polarization, named as realistic $NN$ interaction, while the other one is extracted from the properties of finite nuclei and infinite nuclear matter~\cite{bender03,stone07,ring96,vretenar05,meng06,niksic11}. The latter is called as the effective $NN$ interaction in terms of the nuclear medium effects.

A reasonable $NN$ force model should not only describe the interacting behaviors of nucleons, i.e., the strong repulsion at short range, the large attraction at intermediate range, and the small tail at long range, but also satisfy several basic symmetries, such as, rotation invariance in space, translation invariance, space reflection invariance, and so on. Therefore, at the early stage, the $NN$ forces were constructed in the coordinate space in terms of the angular momentum operators which accord with the symmetry requirements, such as Hamada-Johnston potential~\cite{hamada62} and Reid68 potential~\cite{reid68} in the non-relativistic framework. On the other hand, the one-boson-exchange (OBE) potentials were proposed based on the development of quantum field theory, where the nucleon interacts with each other by exchanging several mesons whose masses are below $1$ GeV~\cite{erkelenz74}. With the great achievements of OBE potentials, more degrees of freedom, like multi-meson exchange and $\Delta$ isobar were introduced, which generated the Bonn full models~\cite{machleidt87,machleidt89}.

In 1990s, the effects of charge independence breaking (CIB) and charge symmetry breaking (CSB) were introduced in nuclear force due to the detailed analysis from the thousands of $NN$ scattering data. The high-precision $NN$ forces were constructed such as, Reid93, Nijmegen 93, Nijmegen I, Nijmegen II, and AV18 potentials~\cite{stoks94,wiringa95}. The chiral perturbation theory was also applied to derive the $NN$ interaction firstly proposed by Weinberg~\cite{weinberg90,weinberg91,weinberg92}. The chiral $NN$ potentials have been built up to the fifth-order expansion until now~\cite{ordonez94,ordonez96,epelbaum98,epelbaum00,entem03,epelbaum05,entem15,epelbaum15a,epelbaum15b,entem17,reinert17}.  In 2000, a covariant charge-dependent OBE (CD-Bonn) potential was proposed by Machleidt as a very high-precision $NN$ interaction, which can describe the $NN$ scattering data very well, with $\chi^2/\text{datum}\sim 1$~\cite{machleidt01}. There are $\omega,~\rho, ~\pi$ mesons and two scalar mesons $\sigma_1$ and $\sigma_2$ in CD-Bonn potential, which was widely applied to study the properties of nuclear systems, from light nuclei to heavy nuclei and infinite nuclear matter.      

The pion as the first discovered meson mainly determines the behavior of $NN$ interaction at long-range region. It also denotes one of the most crucial characters of QCD theory, chiral symmetry, as a Goldstone boson. However, the coupling between pion and nucleon has two possible ways, namely the pseudoscalar (PS) or pseudovector (PV) coupling. Actually, the on-shell amplitudes of one-pion exchange from the PS and PV $\pi NN$ couplings are identical. Caia {\it{et al.}} also examined the differences between PS and PV couplings of pion and $\eta$ meson in $NN$ potentials in terms of two kinds of scattering equations, i.e., Lippmann-Schwinger equation and Thompson equation~\cite{caia02}. It was found that the differences of phase shifts and binding properties for two-nucleon system were very small between PS and PV couplings in these two approaches. The largest difference between these two couplings appeared in the mixing parameter $\varepsilon_1$ at $J=1$ channel. This mixing parameter is strongly related to the strength of tensor component in $NN$ interaction. Therefore, the $D$-state probability, quadrupole moment and asymptotic $S$-state amplitude of deuteron have a few distinctions between the PS and PV couplings.

On the other hand, in the calculations of pion-nucleon scattering, the unitarity and analytic continuation of $\pi N$ to the $\pi\pi\rightleftharpoons N\bar N$ and so on, the PV coupling was preferred~\cite{lacombe75,jackson75}. In the chiral perturbation theory, the coupling between pion and nucleon was also taken as the PV type in low energy region to analyze the pion electroproduction and photoproduction~\cite{drechsel92,drechsel99}. Furthermore, when the $NN$ potentials with PS coupling were applied to the relativistic nuclear many-body methods, a spurious strong attraction was generated due to its strong coupling to negative energy states, which is absent in non-relativistic framework, while the PV type suppresses its coupling to the antinucleons since its matrix element between the antinucleon and nucleon vanishes in the on-shell scattering~\cite{fuchs98}. For example, the PV couplings were adopted in Bonn A, B, C potentials, which were successfully used in the relativistic Bruckner-Hartree-Fock method and led to relatively reasonable saturation properties of nuclear matter~\cite{brockmann90}.

In this work, we would like to develop a revised version of CD-Bonn potential with PV coupling for pion, which can be applied to the relativistic nuclear many-body methods. The theoretical framework of CD-Bonn potential is kept except using the PV coupling instead of PS coupling between pion and nucleon. As an attempt, the coupling constants and the cut-off momenta in form factors for various mesons will be determined by fitting the phase shifts from Nijmegen PWA in present stage. The paper is arranged as follows: in Sec.~\ref{sec.model}, the necessary formulas about the OBE potential and $NN$ scattering are presented. In Sec.~\ref{sec.results}, the CD-Bonn potentials with PV coupling are shown and the properties of two-nucleon scattering and binding states from these potentials are presented. Summary and outlook are given in Sec.~\ref{sec.sum}. 

\section{The CD-Bonn potentials with  pseudovector pion-nucleon coupling }\label{sec.model}

In conventional OBE potential, there are six mesons whose masses are below $1$ GeV, i.e., $\sigma,~\omega,~\rho,~\pi,~\eta,~\delta$ mesons~\cite{machleidt87,machleidt89}. In CD-Bonn potential, two heavier mesons, $\delta$ and $\eta$ were not considered~\cite{machleidt01}. Furthermore, to better simulate the broad contribution from the $2\pi+\pi\rho$ exchange in the intermediate range between two nucleons, two scalar mesons $\sigma_1$ and $\sigma_2$ were included. The Lagrangians which describe mesons-nucleon couplings in CD-Bonn potential are given as:
\begin{align}
\label{NNS}
\mathcal{L}_{\sigma NN}=& -g_\sigma \bar{\psi}\psi\varphi^{(\sigma)},\\
\label{NNO}
\mathcal{L}_{\omega NN}=& -g_\omega \bar{\psi}\gamma^\mu \psi \varphi^{(\omega)}_\mu,\\
\label{NNR}
\mathcal{L}_{\rho NN}  = & -g_\rho \bar{\psi}\gamma^\mu \vec{\tau}\psi\cdot\vec{\varphi}_\mu^{(\rho)} -
\frac{f_\rho}{4M_p} \bar{\psi}\sigma^{\mu\nu}\vec{\tau}\psi\cdot[\partial_\mu \vec{\varphi}_\nu^{(\rho)} -
\partial_\nu \vec{\varphi}_\mu^{(\rho)}],
\end{align}
for $\sigma,~\omega$, and $\rho$ mesons. Here $\psi$ denotes the nucleon field and $\varphi$ represents the meson field. The tensor coupling between $\omega$ meson and nucleon is neglected due to its small strength. To apply the CD-Bonn potential to study the nuclear many-body system in the relativistic framework, the pseudovector (PV) coupling between pion and nucleon is taken in this work,
\begin{eqnarray}
\label{NNP-pv}
\mathcal{L}_{\pi NN}^{(\text{pv})} &= -\frac{f_\pi}{m_{\pi}}\bar{\psi}\gamma^5\gamma^\mu\vec{\tau}\psi\cdot\partial_\mu\vec{\varphi}^{(\pi)},
\end{eqnarray}
where the coupling constant $f_\pi$ is related to the pseudoscalar (PS) coupling constant $g_\pi$, by the on-shell-equivalent relation~\cite{brockmann90} 
\begin{equation}\label{pi-equiv}
\frac{g_\pi}{M_1+M_2} =\frac{f_\pi}{m_\pi}.
\end{equation}
Here $M_1$ and $M_2$ are the masses of two nucleons, respectively, and $m_\pi$ is the pion mass. 

The contribution from each meson $\alpha$ to the $NN$ interaction is expressed analytically by the scattering amplitude from the quantum field theory~\cite{machleidt89}, 
\begin{equation}\label{NN-Amp}
i\bar{V}_\alpha(\mathbf{q}',\mathbf{q}) = \bar{u}_1(\mathbf{q}')\Gamma_1^{\alpha}u_1(\mathbf{q})\frac{P_\alpha}{\mathbf{k}^2 +m_\alpha^2} \bar{u}_2(\mathbf{-q}')\Gamma_2^{\alpha}u_2(-\mathbf{q}),~~(\mathbf{k} = \mathbf{q}'-\mathbf{q}),
\end{equation}
where $\mathbf{q}'$ and $\mathbf{q}$ are the relative momenta of two nucleons for in- and out-scattering states in center-of-mass (CM) framework. The vertex $\Gamma_{i}^\alpha~(i=1,2)$ and the meson propagator $P_\alpha/(\mathbf{k}^2+m_\alpha^2)$ can be directly generated from Lagrangians in Eqs. (\ref{NNS}) and (\ref{NNP-pv}). $u(\mathbf{q})$ is the Dirac spinor of nucleon. 
The static approximation is adopted in the denominator of meson propagator to obtain an energy-independent $NN$ potential. The explicit expressions for involved mesons are shown in the appendix \ref{sec.app1}.

In addition, a form factor $\mathcal{F}_\alpha(\mathbf{k}^2)$ should be introduced to treat the finite size of nucleon. There are many choices for $\mathcal{F}_\alpha(\mathbf{k}^2)$, like monopole form, dipole form, exponential form, and so on. In CD-Bonn potential, the monopole form factor is adopted at each vertex between meson and nucleon,
\begin{equation}\label{formfactor}
\mathcal{F}_\alpha(\mathbf{k}^2) = \frac{\Lambda_\alpha^2-m_\alpha^2}{\Lambda_\alpha^2+\mathbf{k}^2},
\end{equation}
where $m_\alpha$ is the mass of meson and $\Lambda_\alpha$ the corresponding cut-off momentum. 

The Bethe-Salpeter (BS) equation is used to describe the $NN$ scattering in relativistic framework. However, it is very difficult to solve this four-dimensional integral equation. A three-dimensional reduction should be done to achieve the numerical results. There are many schemes to do such reduction, such as Blankenbecler-Sugar (BbS) choice~\cite{blankenbecler66}, Thompson choice~\cite{thompson70}, Kadyshevsky choice~\cite{kadyshevsky68} and so on. The BbS choice was taken in the original CD-Bonn potential. The Bonn A, B, C potentials adopted the Thompson choice. It must be emphasized that we just want to obtain a revised CD-Bonn potential with PV coupling and compare it with the old version now. Therefore, the BbS choice is still used in this work.

When the BbS choice is taken in the propagator part, the BS equation in the two-nucleon CM frame is reduced to~\cite{machleidt01},  
\begin{equation}\label{Bbs-eq}
\bar {T}(\mathbf{q}',\mathbf{q}) = \bar V(\mathbf{q}',\mathbf{q})+
\int \frac{d^3 \mathbf{k}}{(2\pi)^3} ~\bar{V}(\mathbf{q}',\mathbf{k})\frac{M^2}{E_\mathbf{k}}\frac{1}{\mathbf{q}^2-\mathbf{k}^2+i\epsilon}\bar T(\mathbf{k},\mathbf{q})
\end{equation}
After taking minimal relativity and the form factors, the complete interaction is constructed by
\begin{equation}\label{NN-potential}
V(\mathbf{q}',\mathbf{q}) = \sum_{\text{All Mesons}}\mathcal{F}^2_\alpha(\mathbf{k}^2)
 \sqrt{\frac{M}{E'}}\bar{V}_\alpha(\mathbf{q}',\mathbf{q})
 \sqrt{\frac{M}{E}}
\end{equation}
where $E=\sqrt{\mathbf{q}^2 +M^2}$,  $E_\mathbf{k} = \sqrt{\mathbf{k}^2 +M^2}$, 
$E'=\sqrt{\mathbf{q}'^2 +M^2}$ are the starting, intermediate and final energies, respectively.
Then the scattering equation of two-nucleon is given as a three-dimensional BbS equation
\begin{equation}\label{LS-eq}
T(\mathbf{q}',\mathbf{q}) = V(\mathbf{q}',\mathbf{q})+\int\frac{d^3 \mathbf{k}}{(2\pi)^3} ~V(\mathbf{q}',\mathbf{k})\frac{M}{\mathbf{q}^2-\mathbf{k}^2+i\epsilon}T(\mathbf{k},\mathbf{q}).
\end{equation}
This equation has the similar form with Lippmann-Schwinger equation. The phase shifts of $NN$ scattering can be obtained from on-shell $T$-matrix. The mathematical details are presented in Appendix~\ref{sec.app1} as well.

\section{Results and discussions}\label{sec.results}
\subsection{The pion effects in CD-Bonn potentials}
\begin{figure}[h]
	\centering
	\includegraphics[scale=0.75]{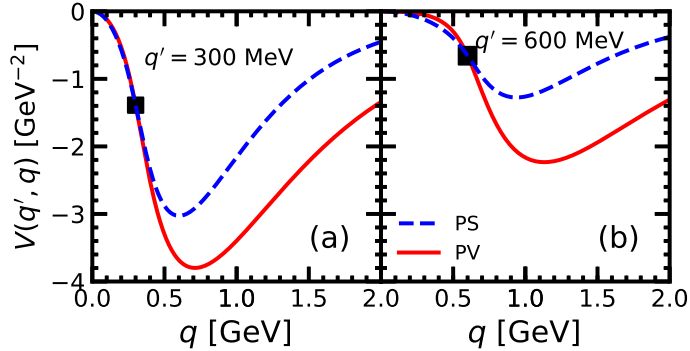}\\
	\caption{\small{Half on-shell matrix elements of OPE potential at $^3S_1$-$^3D_1$ coupled channel. 
			In panel (a),  the on-shell point is fixed at $q'=300$ MeV, and in panel (b), the on-shell  point is fixed at $q'=600$ MeV. The dashed line represents the PS coupling, while the solid curve denotes the PV type.}}\label{pion}
\end{figure}

The on-shell matrix elements of one-pion-exchange (OPE) potentials within PS and PV couplings are identical, while their off-shell behaviors are completely difference, especially in the high momentum region. The half on-shell matrix elements of OPE potentials, $V(q,q')$ from PS and PV couplings at $^3S_1$-$^3D_1$ channel, with fixed on-shell momenta $q'=300$ MeV and $q'=600$ MeV are shown in Fig.~\ref{pion} (a) and Fig.~\ref{pion} (b), respectively. The cut-off momentum $\Lambda_\pi=1720$~MeV and the pion-nucleon coupling constant $g_\pi^2/4\pi = 13.6$ are used here. The matrix elements of OPE potential at $^3S_1$-$^3D_1$ channel are generated by its tensor component. It is found that the PV coupling provides more attractive contribution at higher momentum region comparing to the PS case, while they have the similar strength of matrix elements before on-shell momenta.

The PV coupling constant $f_\pi$ is related to PS coupling constant $g_\pi$ as shown in Eq. (\ref{pi-equiv}). Traditionally, the $\pi N$ coupling constants are expressed as $g_\pi$ for both PS coupling and PV coupling. To consider the charge symmetry breaking effect, $g_\pi$ in neutron-neutron ($nn$), proton-proton ($pp$), and neutron-proton ($np$) systems should be distinguished. Its value in $np$ and $nn$ systems can be expressed by $g_\pi(pp)$ with following relations~\cite{machleidt01},
\begin{eqnarray}\label{P-const}
\frac{g_\pi^2(np)}{4\pi} &=& \left(\frac{M_p+M_n}{2M_p}\right)^2\frac{g_\pi^2(pp)}{4\pi},\\\nonumber
\frac{g_\pi^2(nn)}{4\pi} &=& \left(\frac{M_n}{M_p}\right)^2\frac{g_\pi^2(pp)}{4\pi}.
\end{eqnarray}

In this work, the coupling constants and cut-off momenta of $\omega$ and $\rho$ mesons are directly taken from the original CD-Bonn potential with PS coupling for all channels.  The effect of tensor force in nuclear many-body system is a very hot topic in recent research. The strength of OPE potential determines the magnitude of tensor force in $NN$ interaction directly. Therefore, we would like to choose three pion-nucleon coupling constants to produce  the different tensor components following the idea of Bonn A, B, C potentials~\cite{brockmann90}. The corresponding coupling constants and cut-off momenta are listed in Table \ref{meson-params}. To reduce the uncertainties of coupling constants as much as possible, there is a constraint between $g_\pi$ and $\Lambda_\pi$ at free space, $\mathcal{F}^2_\pi(\mathbf{k}^2=0)\cdot g_\pi^2/4\pi=13.42$. For convenience, in the following discussion, these three potentials with PV coupling are named as pvCD-Bonn A, B, C potentials, respectively. 
\begin{tiny}
	\begin{table}[h]
		\centering
		\caption{\small{The coupling constants and cut-off momenta of pion and $\omega,~\rho$ mesons in pvCD-Bonn A, B, C potentials.}}\label{meson-params}
		\begin{tabular}{c|ccc|ccc|ccc}
			\hline
			\hline
			&\multicolumn{3}{c|}{ A}&\multicolumn{3}{c|}{B}&\multicolumn{3}{c}{C}\\
			$m_a$~[MeV]&~$g_a^2/4\pi$~&~$f_a/g_a$~&~$\Lambda_a$[GeV]~&~$g_a^2/4\pi$~&~$f_a/g_a$&~$~\Lambda_a$[GeV]~&~$g_a^2/4\pi$~&~$f_a/g_a$&~$~\Lambda_a$[GeV]\\
			\hline
			$\pi^0$ (139.57)              &  13.9     &      &   1.12  & 13.7       &      &  1.50& 13.6       &      &  1.72 \\
			$\pi^\pm$(134.98)             &  13.9     &      &   1.12  & 13.7       &      &  1.50& 13.6       &      &  1.72 \\
			$\rho^{0},\rho^\pm$ (770)     &  0.84     &  6.1 &   1.31  & 0.84       &  6.1 &  1.31& 0.84       &   6.1&  1.31 \\
			$\omega$(782)                 &  20       &      &   1.50  &  20        &      &  1.50& 20         &      &  1.50   \\
			\hline
			\hline
		\end{tabular}
	\end{table}
\end{tiny}

\begin{figure}[h]
	\centering
	\includegraphics[scale=0.75]{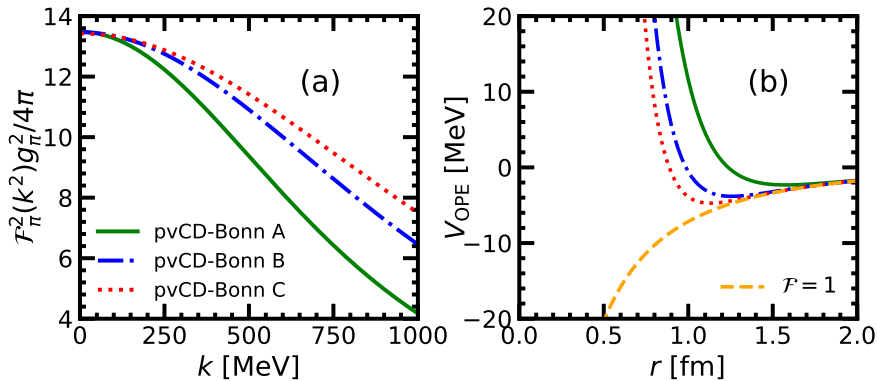}\\
	\caption{\small{The effective $\pi NN$ coupling strengths of pvCD-Bonn A, B, C potentials as functions of the transferred momentum (a) and the corresponding isopin-singlet, spin-triplet OPE potentials in coordinate space (b).}}\label{pion-coupling}
\end{figure}

The form factor can suppress the OPE potential at high momentum region largely. In Fig.~\ref{pion-coupling} (a), the products of $g^2_\pi/4\pi$ and $\mathcal{F}^2_\pi(k^2)$, which can be regarded as effective coupling constants, for pvCD-Bonn A, B, C potentials are given as functions of momenta. Due to the influence of form factor, the effective coupling constants decrease rapidly with the momentum increasing. The $\pi N$ coupling constant in pvCD-Bonn A is the largest one, however its pion contribution is the smallest in the three pvCD-Bonn potentials due to its minimum cut-off momentum. Its magnitude at $k=1000$ MeV is nearly half of the one from pvCD-Bonn C potential.

In Fig.~\ref{pion-coupling} (b) , the local OPE potentials from pvCD-Bonn A, B, C in coordinate space are plotted at isospin-singlet and spin-triplet channel, i. e. $T=0,~S=1$. To show the influence of form factor, a free OPE potential is also given to be compared. The form factor mainly plays its role at high momentum region. Correspondingly, it exhibits the cut-off effect at short-range part in coordinate space. When the relative distance between two nucleons is larger than $1.5$ fm, the OPE potentials with and without form factor are almost identical. In the short-range region, the form factor changes the OPE potentials significantly. 

\subsection{The phase shifts of $NN$ scattering from pvCD-Bonn potentials}
To describe the CSB of $NN$ interaction precisely, the coupling constants and cut-off momenta of $\sigma_1$ and $\sigma_2$ mesons are fitted to reproduce the phase shifts of $NN$ scattering analyzed by Nijmegen group in 1993, i.e., Nijmegen PWA~\cite{stoks94} at each partial wave with $J\leq 4$ following the scheme of Machleidt~\cite{machleidt01}. The $NN$ laboratory energy is up to $300$ MeV. The BbS equation is solved by matrix inversion method~\cite{haftel70} through discretizing the integration with Gauss-Laguerre quadrature. The free parameters are looked for by a  numerical minimization Fortran program, MINUIT.

The fitting optimal function is defined as
\begin{equation}
\chi^2=\frac{1}{N}\sum^{N}\left(\frac{\delta_\text{OBE}-\delta_\text{NM}}{\delta_\text{NM}}\right)^2,
\end{equation}
where $\delta_\text{OBE}$ are the phase shifts predicted by pvCD-Bonn A, B, C potentials and $\delta_\text{NM}$ are the data from Nijmegen PWA. Consequently, $\chi^2$ used in this work only reflects the deviations of phase shifts between theoretical calculations and Nijmegen PWA. The fitting energies cover the $NN$ elastic laboratory energy region $0-350$ MeV. We take $E_\text{lab}=1,~5,~10,~25,~50,~100,~150,~200,~250,~300,~350$ MeV  in each channel. The Nijmegen PWA data are obtained from http://nn-online.org.

As shown in the original CD-Bonn potential, at isospin $T=1$ channel, $pp, nn$ and $np$ interactions are not independent due to CSB and CIB. Once one of these three interactions is determined, the other two channels will be fixed. Therefore, we firstly fit the $pp$ interaction, whose scattering data have been measured most accurately until now. The Coulomb interaction should be included in $pp$ scattering as a long-range potential. The present asymptotic wave functions are related to regular and irregular Coulomb functions. The detailed technologies were shown in the appendix part of Ref.~\cite{machleidt01}. 

\begin{figure}[h]
	\centering
	\includegraphics[scale=0.55]{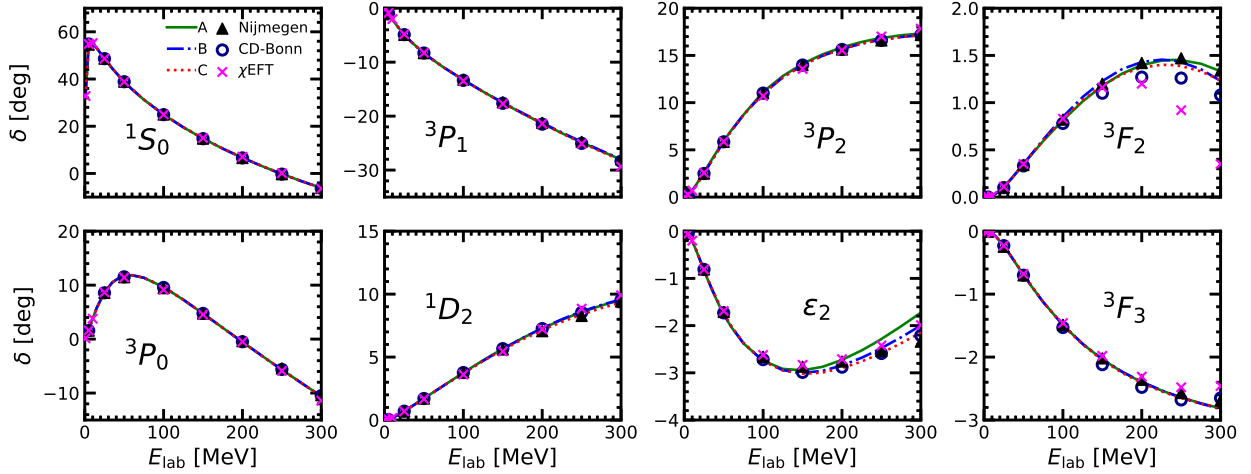}\\
	\caption{\small{The $pp$ phase shifts and mixing parameters as functions of laboratory energies, $E_\text{lab}$ ($J\le 3$). The triangles, open circles, and crosses represent the phase shifts predictions from Nijmegen PWA, CD-Bonn potential, chiral N$^4$LO potential by Entem {\it et al.}, respectively.}}
	\label{pp-PhaseShift}
\end{figure}

With different strengths of OPE potentials, the pvCD-Bonn A, B, C interactions are obtained. Their phase shifts at each partial wave with total angular momentum $J\leq 3$ and the mixing parameters $\varepsilon_2$ are given in Fig.~\ref{pp-PhaseShift}. The fitting data in Nijmegen PWA are also shown as solid triangle, while the phase shifts from CD-Bonn potential with PS coupling are presented as open circles and those from the latest chiral N$^4$LO potential with $\Lambda=500$ MeV by Entem {\it et al.}~\cite{entem17} are given as crosses. It is found that all data from Nijmegen PWA can be described very well and are consistent among these three potentials. There are slightly differences among pvCD-Bonn A, B, C potentials at the phase shifts of $^3F_2$ channel and the mixing parameter, $\varepsilon_2$ at high input $E_\text{lab}$. The $^3F_2$ and $^3P_2$ channels couple together due to the tensor operator. The mixing parameter, $\varepsilon_2$, presents the strength of $NN$ tensor interaction at $J=2$. Therefore, these differences are easily understood due to the different tensor components in pvCD-Bonn A, B, C interactions. The phase shifts and mixing parameters generated by pvCD-Bonn C interaction are highly consistent with those from Nijmegen PWA and CD-Bonn potential at each partial wave, where the amplitudes of OPE potential are largest. Actually, the similar results were revealed in Bonn A, B, C potentials.Furthermore, it can be found that the phase shifts of $^3F_2$ channel from the chiral potential are obvious deviations from other data. This is because that the phase shifts of $^3F_2$ channel should be correctly described by sixth-order expansion in chiral potential~\cite{reinert17}.

\begin{figure}[h]
	\centering
	\includegraphics[scale=0.55]{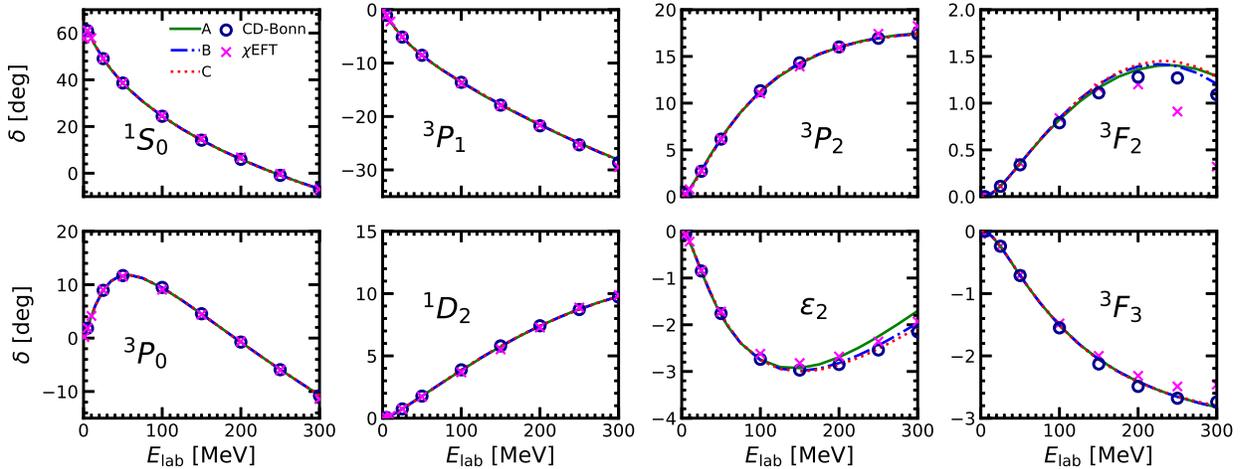}\\
	\caption{\small{The $nn$ phase shifts and mixing parameters from pvCD-Bonn A, B, C potentials at each partial wave ($J\le 3$). The open circles and crosses denote the results from CD-Bonn potential and chiral N$^4$LO potential by Entem {\it et al.}}}
	\label{nn-PhaseShift}
\end{figure}

The $nn$ interactions were generated based on the $pp$ interactions, by interchanging the proton mass to neutron mass and fitting the coupling constants of $\sigma_1$ and $\sigma_2$ mesons with the CSB phase shift differences from the nucleon mass splitting on kinematics, OBE diagram, and two-boson exchanges worked out by Machleidt in Ref.~\cite{machleidt01}. The phase shifts and mixing parameters of $nn$ interactions from pvCD-Bonn A, B, C potentials are shown in Fig.~\ref{nn-PhaseShift}. There is no phase shift data about $nn$ scattering in Nijmegen PWA. Therefore, the  phase shifts and mixing parameters from pvCD-Bonn A, B, C potentials are only compared with the results from CD-Bonn potential with PS coupling. They are slightly different at coupled channels, $^3F_2$-$^3P_2$ waves and mixing parameters, $\varepsilon_2$, which are very similar with the $pp$ case. 

\begin{figure}[h]
	\centering
	\includegraphics[scale=0.55]{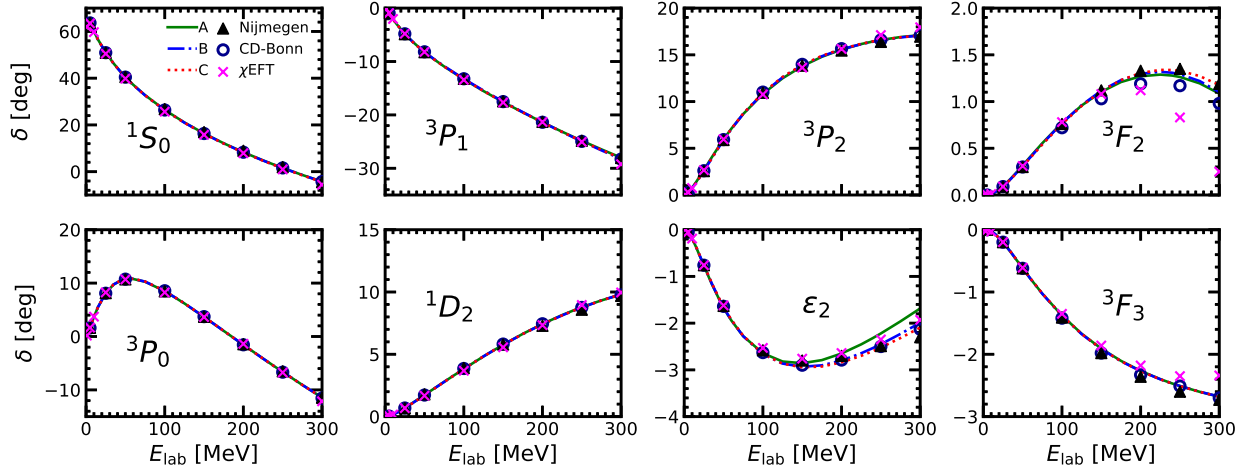}\\
	\caption{\small{The $np~(T=1)$ phase shifts and mixing parameters in different partial waves ($J\le 3$).
			The triangles, open circles, and crosses represent the phase shifts predictions from Nijmegen PWA, CD-Bonn potential, chiral N$^4$LO potential by Entem {\it et al.}, respectively.}}
	\label{np-PhaseShift1}
\end{figure}

For the $np~(T=1)$ potentials, the analogous procedure has been done. The proton mass is replaced by the average mass, $\widetilde M=\sqrt{M_n M_p}$. Furthermore, in the OPE potential, the positive and negative pions are also included in the exchanging between two nucleons, whose masses are slightly different with the $\pi^0$ mass. The coupling constants of $\sigma_1$ and $\sigma_2$ for $np$ potentials are determined by adjusting the CIB phase shift differences between $pp$ and $np$ potentials from nucleon mass splitting, OPE, two-boson exchange, irreducible $\pi\gamma$ exchange, and Coulomb force. The phase shifts and mixing parameters of $np ~(T=1)$ at each partial wave of $J\leq 3$ are presented as functions of laboratory kinetic energies in Fig.~\ref{np-PhaseShift1}. The distinctions about the phase shifts and mixing parameters among pvCD-Bonn A, B, C potentials at $^3F_2$ channel and $\varepsilon_2$ still exist.

\begin{figure}[h]
	\centering
	\includegraphics[scale=0.55]{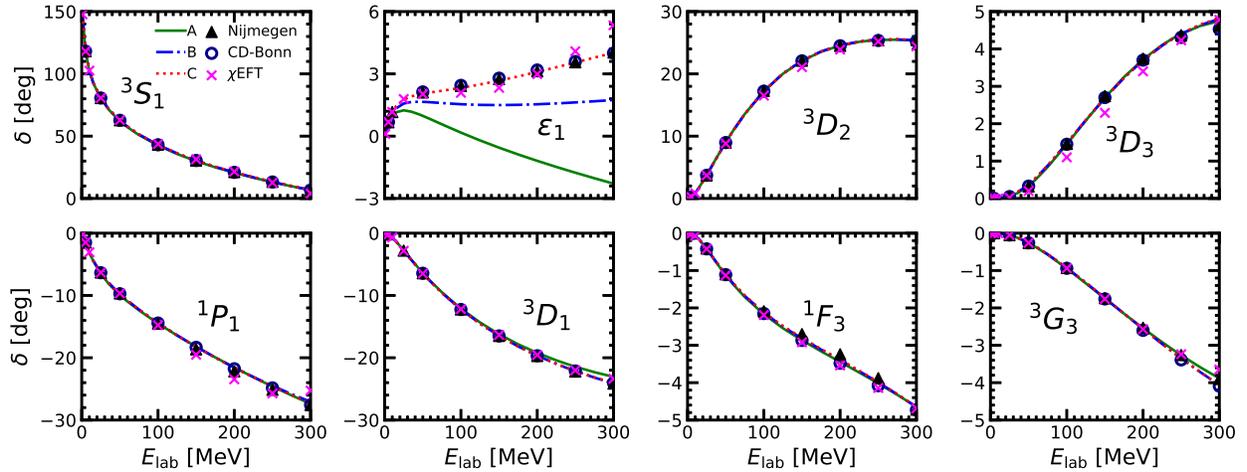}\\
	\caption{\small{The $np~(T=0)$ phase shifts and mixing parameters in different partial waves ($J\le 3$).
			The triangles, open circles, and crosses represent the phase shifts predictions from Nijmegen PWA, CD-Bonn potential, chiral N$^4$LO potential by Entem {\it et al.}, respectively.}}
	\label{np-PhaseShift0}
\end{figure}

For the $np$ potential, there is an other possibility with $T=0$ channel. The corresponding coupling constants of $\sigma_1$ and $\sigma_2$ mesons are directly adjusted to reproduce the phase shifts from Nijmegen PWA. The phase shifts and mixing parameters from pvCD-Bonn A, B, C potentials are plotted in Fig.~\ref{np-PhaseShift0}. The fitting data from Nijmegen PWA and the calculations from CD-Bonn potentials are also shown. The strongest tensor component in $NN$ potential comes from the coupled channels $^3S_1$ and $^3D_1$. Their mixing parameter, $\varepsilon_1$ is very sensitive to the strength of OPE potential. Their values obtained from pvCD-Bonn C potential can completely simulate the results from Nijmegen PWA and CD-Bonn potential, while the $\varepsilon_1$ from pvCD-Bonn A and B potentials have significant differences with the fitting data. The phase shifts at other partial waves from pvCD-Bonn A, B, C potentials are almost identical and describe the results from Nijmegen PWA very well. 

As an example, the coupling constants and masses of $\sigma_1$ and $\sigma_2$ mesons for $pp$, $nn$, $np$ components of pvCD-Bonn C potential up to total angular momentum $J=4$ are listed in Table~\ref{sigma-paramC}. Both the cut-off momenta of $\sigma_1$ and $\sigma_2$ mesons are taken as $\Lambda_{\sigma_1,~\sigma_2}=2500$ MeV. The blank denotes that there is no meson contribution. The detailed values of phase shifts for $pp$, $nn$, $np$ scattering are tabulated in Tables \ref{psppc}-\ref{psnpc0}. For all channels whose total angular momenta are larger than $J=4$, there is only one scalar meson considered. Its mass and coupling constant are $m_{\sigma}=452$ MeV, $g^2_\sigma/4\pi=2.3$ for pvCD-Bonn B, C and $m_{\sigma}=470$ MeV, $g^2_\sigma/4\pi=4.3$ for pvCD-Bonn A, respectively, following the treatments in Ref.~\cite{machleidt01}. The coupling constants and phase shifts of pvCD-Bonn A and B potentials are shown detailed in Appendix A. Finally, the $\chi^2$ of pvCD-Bonn A, B, C potentials are $3.55\times 10^{-3}, ~1.85\times 10^{-3}$ and $2.24\times 10^{-4}$, respectively for $pp$ interactions and $8.06\times 10^{-2}, ~1.03\times 10^{-2}$ and $2.65\times 10^{-4}$ for $np(T=0)$ interaction.

\begin{table}[htbp]
	\centering
	\caption{ Parameters of $\sigma_1$ and $\sigma_2$ in pvCD-Bonn C. Blanks indicate the corresponding parameters taken as zero. Meson masses are in unit MeV.}\label{sigma-paramC}
	\begin{tabular}{c|cc|cc|cc|cc}
		\hline
		\hline
		&$g^2_{\sigma_1}/4\pi$&$g^2_{\sigma_2}/4\pi$&$g^2_{\sigma_1}/4\pi$&$g^2_{\sigma_2}/4\pi$ &$g^2_{\sigma_1}/4\pi$&$g^2_{\sigma_2}/4\pi$&$m_{\sigma_1}$&$m_{\sigma_2}$ \\
		\hline
		$^1S_0$&   5.17     &  4.00    &  4.87    &  10.40   & 5.19     &  3.92    &  470  &  1225 \\
		$^3P_0$&   4.32     &  38.04   &  4.32    &  27.08   &  4.34    & 36.15    &  500  &  1225 \\
		$^1P_1$&            &          &  1.43    &  73.51   &          &          &  400  &  1225 \\
		$^3P_1$&  2.29      &  70.04   &  2.26    &  69.76   &  2.29    &  73.00   &  424  &  1225 \\
		$^3D_1$&            &          &  2.56    &  12.43   &          &          &  452  &  793  \\
		$^3S_1$&            &          &  2.29    &   1.42   &          &          &  452  &  793  \\
		$^1D_2$&  2.20      &  202.55  &   2.21   &  201.53  & 2.25     &  198.42  &  400  & 1225  \\
		$^3D_2$&            &          &   0.67   &  59.26   &          &          &  350  & 1225  \\
		$^3F_2$&  1.80      &  31.65   &   1.68   &  35.54   & 1.78    &  32.32  &  424  &  793  \\
		$^3P_2$&  3.29      &  29.33   &   3.28   &  29.48   & 3.29     & 29.32   &  452  & 1225  \\
		$^1F_3$&            &          &   0.90   &          &          &          &  400  &       \\
		$^3F_3$&  2.70      &  45.81   &   2.82   &  43.50   &  2.70    &  45.81   &  452  &  793  \\
		$^3G_3$&            &          &   1.68   &          &          &          &  350  &       \\
		$^3D_3$&            &          &   1.52   &          &          &          &  350  &       \\
		$^1G_4$&  3.90      &          &   3.90   &          &  3.90    &          &  452  &       \\
		$^3G_4$&            &          &   3.90   &          &          &          &  452  &       \\
		$^3H_4$&  3.36      &          &   3.36   &          &  3.36    &          &  452  &       \\
		$^3F_4$&  3.80      &          &   3.80   &          &  3.80    &          &  452  &       \\
		\hline
		&\multicolumn{2}{c|}{$pp$}&\multicolumn{2}{c|}{$np$}&\multicolumn{2}{c|}{$nn$}&\multicolumn{2}{c}{}\\
		\hline
		\hline
	\end{tabular}
\end{table}

\begin{table}
	\centering
	\caption{$pp$ phase shifts in different partial waves, predicted by pvCD-Bonn C. }\label{psppc}
	\setlength{\tabcolsep}{2.5mm}{
		\begin{tabular}{ccccccccccc}
			\hline
			\hline
			$T_\text{lab}$(MeV)&$^1S_0$&$^3P_0$&$^3P_1$&$^1D_2$&$^3P_2$&$\varepsilon_2$&$^3F_2$&$^3F_3$&$^1G_4$&$^3F_4$\\
			\hline
			1  & 32.77 & 0.13  & -0.08 &  0.00 &  0.01 &  0.00 &  0.00 & -0.00 &  0.00 & 0.00  \\
			5  & 54.85 & 1.59  & -0.90 &  0.04 &  0.22 & -0.05 &  0.00 & -0.01 &  0.00 & 0.00  \\    
			10  & 55.22 & 3.75  & -2.05 &  0.17 &  0.66 & -0.20 &  0.01 & -0.03 &  0.00 & 0.00  \\
			25  & 48.69 & 8.68  & -4.90 &  0.70 &  2.50 & -0.82 &  0.11 & -0.23 &  0.04 & 0.02  \\
			50  & 38.94 & 11.72 & -8.30 &  1.70 &  5.83 & -1.73 &  0.34 & -0.69 &  0.15 & 0.12  \\
			100 & 24.95 & 9.62  & -13.31&  3.74 & 10.94 & -2.75 &  0.83 & -1.49 &  0.41 & 0.51  \\
			150 & 14.76 & 4.72  & -17.49&  5.62 & 13.95 & -3.02 &  1.20 & -2.05 &  0.68 & 1.05  \\
			200 &  6.59 &-0.51  & -21.26&  7.23 & 15.66 & -2.90 &  1.41 & -2.41 &  0.94 & 1.64  \\
			250 & -0.30 &-5.61  & -24.73&  8.55 & 16.62 & -2.57 &  1.44 & -2.65 &  1.21 & 2.21  \\
			300 & -6.27 &-10.49 & -27.93&  9.60 & 17.11 & -2.13 &  1.29 & -2.81 &  1.48 & 2.71  \\
			\hline
			\hline
	\end{tabular}}
	
\end{table}

\begin{table}
	\centering
	\caption{$nn$ phase shifts in different partial waves, predicted by pvCD-Bonn C. }\label{psnnc}
	\setlength{\tabcolsep}{2.5mm}{
		\begin{tabular}{ccccccccccc}
			\hline
			\hline
			$T_\text{lab}$(MeV)&$^1S_0$&$^3P_0$&$^3P_1$&$^1D_2$&$^3P_2$&$\varepsilon_2$&$^3F_2$&$^3F_3$&$^1G_4$&$^3F_4$\\
			\hline
			1  & 57.40 & 0.21  & -0.12 &  0.00 &  0.02 &  0.00 & 0.00 & -0.00 &  0.00 & 0.00   \\
			5  & 60.91 & 1.86  & -1.04 &  0.05 &  0.27 & -0.06 & 0.00 & -0.01 &  0.00 & 0.00   \\    
			10  & 57.76 & 4.11  & -2.24 &  0.18 &  0.76 & -0.22 & 0.01 & -0.04 &  0.00 & 0.00   \\
			25  & 49.07 & 8.98  & -5.13 &  0.74 &  2.71 & -0.85 & 0.11 & -0.24 &  0.04 & 0.02   \\
			50  & 38.64 & 11.73 & -8.54 &  1.77 &  6.14 & -1.76 & 0.35 & -0.70 &  0.16 & 0.12   \\
			100 & 24.40 & 9.45  & -13.55&  3.86 &  11.29& -2.75 & 0.84 & -1.51 &  0.42 & 0.52   \\
			150 & 14.13 & 4.51  & -17.74&  5.78 &  14.30& -3.00 & 1.20 & -2.06 &  0.68 & 1.07   \\
			200 & 5.94  &-0.78  & -21.50&  7.41 &  15.99& -2.86 & 1.39 & -2.42 &  0.95 & 1.70  \\
			250 & -0.95 &-5.88  & -24.96&  8.73 &  16.93& -2.51 & 1.39 & -2.65 &  1.23 & 2.24  \\
			300 & -6.93 &-10.69 & -28.14&  9.77 &  17.40& -2.06 & 1.21 & -2.80 &  1.50 & 2.74  \\
			\hline
			\hline
	\end{tabular}}
	
\end{table}

\begin{table}
	\centering
	\caption{$np~(T=1)$ phase shifts in different partial waves, predicted by pvCD-Bonn C. }\label{psnpc1}
	\setlength{\tabcolsep}{2.5mm}{
		\begin{tabular}{ccccccccccc}
			\hline
			\hline
			$T_\text{lab}$(MeV)&$^1S_0$&$^3P_0$&$^3P_1$&$^1D_2$&$^3P_2$&$\varepsilon_2$&$^3F_2$&$^3F_3$&$^1G_4$&$^3F_4$\\
			\hline
			1  & 62.10 & 0.18  & -0.11 &  0.00 &  0.02 &  0.00 & 0.00 & -0.00 &  0.00 & 0.00  \\
			5  & 63.69 & 1.62  & -0.93 &  0.04 &  0.26 & -0.05 & 0.00 & -0.00 &  0.00 & 0.00  \\    
			10  & 60.03 & 3.64  & -2.05 &  0.16 &  0.72 & -0.18 & 0.01 & -0.03 &  0.00 & 0.00  \\
			25  & 50.95 & 8.14  & -4.85 &  0.68 &  2.58 & -0.76 & 0.09 & -0.20 &  0.03 & 0.02  \\
			50  & 40.45 & 10.76 & -8.25 &  1.71 &  5.89 & -1.65 & 0.31 & -0.62 &  0.13 & 0.11  \\
			100 & 26.32 & 8.54  & -13.33&  3.83 & 10.94 & -2.66 & 0.76 & -1.38 &  0.38 & 0.48  \\
			150 & 16.22 & 3.67  & -17.58&  5.77 & 13.92 & -2.95 & 1.11 & -1.91 &  0.66 & 1.01  \\
			200 & 8.18  &-1.57  & -21.40&  7.42 & 15.63 & -2.84 & 1.31 & -2.27 &  0.93 & 1.60  \\
			250 & 1.44  &-6.65  & -24.91&  8.76 & 16.59 & -2.52 & 1.33 & -2.51 &  1.21 & 2.16  \\
			300 &-4.41  &-11.43 & -28.14&  9.82 & 17.08 & -2.09 & 1.18 & -2.68 &  1.49 & 2.65  \\
			\hline
			\hline
	\end{tabular}}
	
\end{table}
\begin{table}
	\centering
	\caption{$np~(T=0)$ phase shifts in different partial waves, predicted by pvCD-Bonn C. }\label{psnpc0}
	\setlength{\tabcolsep}{2.5mm}{
		\begin{tabular}{ccccccccccc}
			\hline
			\hline
			$T_\text{lab}$(MeV)&$^1P_1$&$^3S_1$&$\varepsilon_1$&$^3D_1$&$^3D_2$&$^1F_1$&$^3D_3$&$\varepsilon_3$&$^3F_3$&$^3G_4$\\
			\hline
			1  & -0.19 &147.76 & 0.10 & -0.01 &  0.01 & 0.00 &  0.00 &  0.00 & 0.00 & 0.00\\
			5  & -1.50 &118.19 & 0.67 & -0.18 &  0.22 &-0.01 &  0.00 &  0.01 & 0.00 & 0.00\\    
			10  & -3.08 &102.61 & 1.15 & -0.68 &  0.85 &-0.07 &  0.01 &  0.08 & 0.00 & 0.01\\
			25  & -6.42 & 80.58 & 1.76 & -2.80 &  3.73 &-0.42 &  0.07 &  0.55 &-0.05 & 0.17\\
			50  & -9.81 & 62.66 & 2.04 & -6.43 &  8.98 &-1.10 &  0.38 &  1.61 &-0.26 & 0.72\\
			100 &-14.43 & 43.02 & 2.33 & -12.24& 17.22 &-2.12 &  1.48 &  3.48 &-0.93 & 2.16\\
			150 &-18.23 & 30.51 & 2.67 & -16.47& 22.08 &-2.80 &  2.69 &  4.83 &-1.74 & 3.63\\
			200 &-21.60 & 21.09 & 3.09 & -19.67& 24.54 &-3.39 &  3.69 &  5.76 &-2.56 & 5.01\\
			250 &-24.60 & 13.46 & 3.55 & -22.15& 25.47 &-3.99 &  4.41 &  6.39 &-3.35 & 6.26\\
			300 &-27.28 &  6.99 & 4.02 & -24.11& 25.47 &-4.65 &  4.82 &  6.82 &-4.07 & 7.38\\
			\hline
			\hline
	\end{tabular}}
\end{table}

\begin{figure}[h]
	\centering
	\includegraphics[scale=0.6]{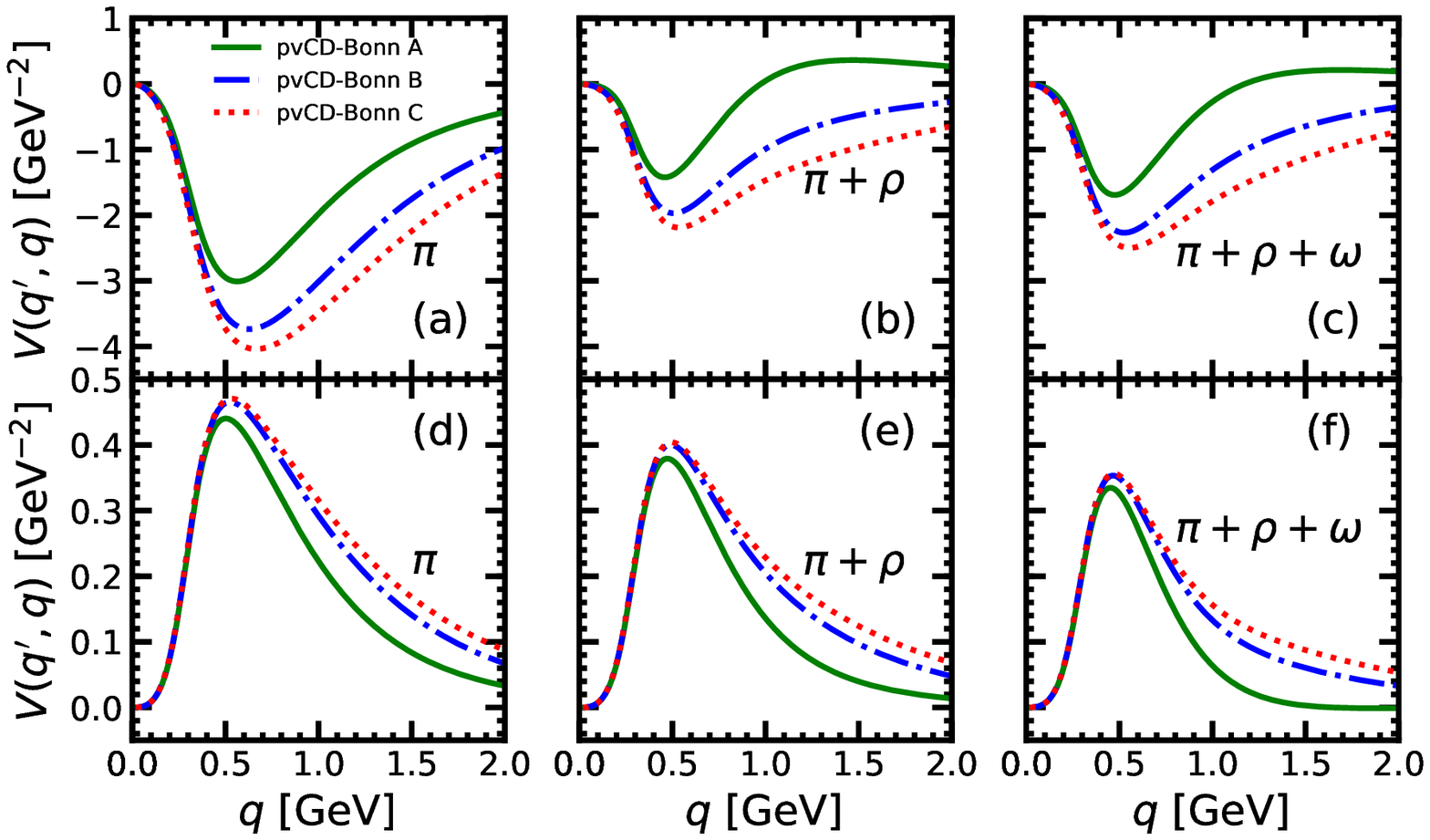}\\
	\caption{\small{The half-off-shell $^3S_1$-$^3D_1$ (panel above) and 
			$^3P_2$-$^3F_2$ (panel behind) potentials. The on-shell momentum is fixed at $q'=265$~MeV.}}
	\label{half-off-shell}
\end{figure}

The tensor force in OBE potential is generated not only from pion but also from $\rho$ and $\omega$ mesons. In Fig.~\ref{half-off-shell}, the half on-shell matrix elements from pion, $\omega$, and $\rho$ mesons in pvCD-Bonn A, B, C potentials at coupled channels, $^3S_1$-$^3D_1$ and $^3P_2$-$^3F_2$ are plotted at on-shell momentum $q'=265$ MeV. The pion provides the largest attractive contribution at $^3S_1$-$^3D_1$ channel, while the tensor component from $\rho$ meson is repulsive. The $\omega$ meson also gives slight attraction. In $^3P_2$-$^3F_2$ channel, the situations for pion and $\rho$ meson are opposite. The $\omega$ meson still provides slightly attractive contributions. 

\subsection{The low-energy scattering parameters and the deuteron}

Once the $NN$ potentials are determined, they can be immediately applied to discuss the lower-energy $NN$ scattering parameters,  such as scattering length, effective range, and the binding properties of the deuteron which have been accurately measured now. At low laboratory energy, the anti-tangent values of phase shifts can be expanded as functions of momenta. The coefficients at first two terms are defined as the scattering length, $a$ and the effective range $r$. Their values at $^1S_0$ and $^3S_1$ channels for $pp$, $nn$, and $np$ scattering from pvCD-Bonn A, B, C potentials are given in Table \ref{lowscat-params}. The corresponding experimental results are also shown in comparison. It can be found that the theoretical calculations from present potentials with PV pion-nucleon coupling are consistent with the experiment results. 

		\begin{table}[h]
		\centering
		\caption{Low-energy $NN$ scattering parameters.
		$a$ represents the scattering length and $r$ the effective range 
		(in unit [fm]), Coulomb effects are excluded in these data.}
		\setlength{\tabcolsep}{0.8mm}{
		\begin{tabular}{ccrcrcrcrcr}
			\hline
			\hline
		             &   &  A      & &  B     & &   C     & &Experiment& & references \\
		   	\hline
		   	$a^N_{pp}$ &   & -17.325 & &-17.292 & & -17.255 & &  & &\\
		   	$r^N_{pp}$ &   & 2.820   & & 2.813  & &  2.809  & &   & & \\
		   	\\
		   	$a_{np}$ &   & -23.711 & &-23.757 & & -23.734 & &-23.74$\pm$0.02 & &\cite{houk71}\\ 
		   	$r_{np}$ &   & 2.649   & &2.649   & &   2.646 & &  2.77$\pm$0.05 & &\cite{houk71}\\
		   	$a_t$    &   & 5.432   & & 5.417  & &   5.417 & & 5.419$\pm$0.007& &\cite{houk71} \\
		   	$r_t$	 &   & 1.773   & &1.753   & &   1.757 & & 1.753$\pm$0.008& &\cite{houk71} \\ 
		   	\\
		   	$a_{nn}$ &   &-18.806  & &-18.744 & & -18.741 & & -18.9$\pm$0.4& &\cite{howell98,gonzalez99}\\
		   	$r_{nn}$ &   & 2.795   & & 2.786  & &  2.784  & & 2.75$\pm$0.11& &\cite{miller90}  \\
		  	\hline
		  	\hline
		\end{tabular}}
		\label{lowscat-params}
		\end{table}

Deuteron is the only bound state of $np$ system. It was found that the wave functions of deuteron should be combined by $S$-state and $D$-state to describe its quadrupole moment reasonably. The solution of deuteron bound state is corresponding to an energy pole in the scattering equation.  Therefore, the wave functions of deuteron can be solved from the scattering equation by introducing the experimental value of deuteron binding energy, $B_d = 2.224575$~MeV~\cite{haftel70}. From these wave functions, the $D/S$-state ratio $\eta$, the asymptotic $S$-state normalization constant $A_S$, the root-mean-square radius of deuteron $r_d$, the quadrupole moment $Q_d$, and the $D$-state probability $P_D$, are predicted by the pvCD-Bonn A, B, C potentials, which are listed in Table \ref{Deutron-results}. Except the $D$-state probability, the other quantities can be measured or extracted from the experiments. Although these three potentials have different tensor components, they provide similar descriptions on the properties of deuteron, which are consistent with the experimental data or empirical values. It should be noted that the quadruple moment in pvCD-Bonn C potentials, $Q_d=0.273$ fm$^2$ is more closed to the experiment data than the one from the CD-Bonn potential with PS coupling, $Q_d=0.270$ fm$^2$. Furthermore, the pvCD-Bonn C potential has the strongest tensor constituent with $P_D=6.06\%$, which has been shown in the Fig.~\ref{half-off-shell}. In Ref. \cite{caia02}, Caia {\it et al.} mentioned that the $P_D$ is $0.4-0.5\%$ higher for PV coupling than PS coupling. In this work, the $P_D$ in pvCD-Bonn C potential is much larger than that in original CD-Bonn potential ($4.85\%$) comparing with this value. The $P_D$ is strongly related to tensor force, which are contributed by pion and $\omega,~\rho$ mesons in CD-Bonn potential. To discuss the difference between PV and PS couplings in this work, their coupling constants and cutoffs in pvCD-Bonn C potential and original CD-Bonn potential are taken the same values, while these parameters in PV and PS couplings were distinguished in the work by Caia {\it et al.} to fit the phase shifts. Therefore, the differences of $P_D$ in pvCD-Bonn C potential and origin CD-Bonn potential are completely generated by the PV and PS couplings.

\begin{table}
\centering
\caption{The deuteron properties predicted by pvCD-Bonn A, B, C potentials. 
		$\eta$ is the $D/S$ ratio,  $A_s$ is the asymptotic $S$-state normalization 
		in fm$^\frac{1}{2}$, $r_d$ is the deuteron matter radius in fm, 
		$Q_d$ is the quadrupole momentum in fm$^2$ and $P_D$ is the $D$-state probability.}
\setlength{\tabcolsep}{2mm}{
\begin{tabular}{ccrcrcrcrcr}
      \hline
      \hline
                       & &     A  & &      B     & &      C     & & Experiment & &  references\\
      \hline
     $\eta$            & & 0.0246 & &    0.0250  & &    0.0253  & &  0.0256(4) & &\cite{rodning90}\\
     $A_S$             & & 0.8895 & &    0.8860  & &    0.8871  & &  0.8883(44)& &\cite{kermode83}  \\
     $r_d$             & & 1.965  & &    1.967   & &    1.967   & &  1.971(6)  & &\cite{martorell95}\\
     $Q_d$             & & 0.261  & &    0.269   & &    0.273   & &  0.2859(3) & &\cite{bishop79}  \\
     $P_D$             & & 4.22\% & &    5.45\%  & &    6.05\%  & &            & & \\
      \hline
      \hline
    \end{tabular}
    }
    \label{Deutron-results}

\end{table}

\begin{figure}[htp]
 \centering
  \includegraphics[scale=0.65]{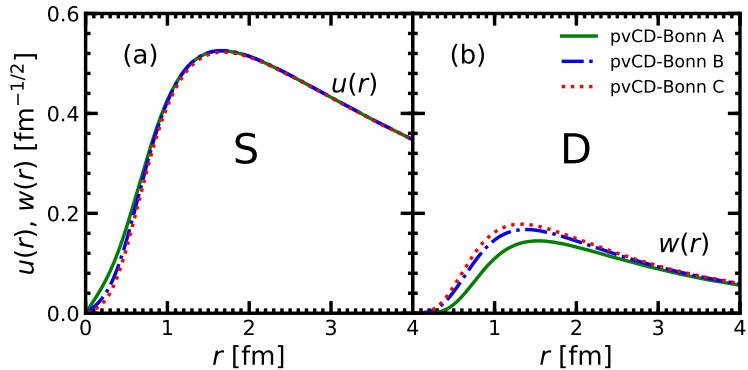}\
  \caption{\small{The normalized deuteron wave functions in configuration space predicted by pvCD-Bonn A, B, C.}}
  \label{Deuteron-wave}
\end{figure}

In Fig.~\ref{Deuteron-wave}, the wave functions of deuteron at $S$-state, $u(r)$ and $D$-state, $w(r)$ from pvCD-Bonn A, B, C potentials are shown in coordinate space. These wave functions were solved firstly in momentum representation with Lippmann-Schwinger equation. They are switched to coordinate space by Fourier transformation. For $S$-state, the wave functions from three potentials are almost identical. There are significant differences in the wave functions of $D$-state at intermediate range between $1-2$ fm. This is just the interaction range of tensor force. The wave function of pvCD-Bonn C potential has the largest amplitude at $D$-state, which produces the strongest $D$-state probability. 

\begin{figure}[htp]
	\centering
	\includegraphics[scale=0.65]{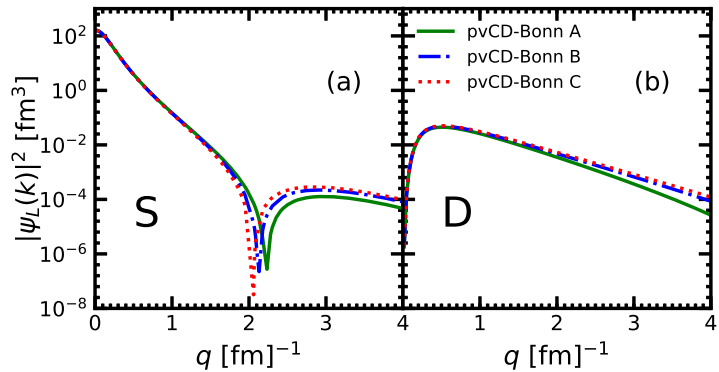}\
	\caption{\small{The squares of deuteron wave functions for $S$ and $D$ states in momentum space predicted from pvCD-Bonn A, B, C.}}
	\label{Deuteronq-wave}
\end{figure}

The squares of deuteron wave functions in momentum space for $L=0$ and $L=2$ states are shown in Fig.~\ref{Deuteronq-wave} obtained with our three potentials. They are very important for calculating the electromagnetic form factors of deuteron~\cite{gilman02}. There are rapid declines for $|\psi(k)|^2$ at $S$ state around $k=2$ fm$^{-1}$, due to the strong repulsion of $NN$ interaction at short distance, which changes the sign of deuteron wave function. Actually, there are also the similar behaviors for deuteron wave functions from other realistic $NN$ interactions~\cite{bogner06,gilman02}. At small momentum regions, these wave functions from pvCD-Bonn A, B, C potentials are very similar, while they have obvious differences at high momentum region, which correspond to the short-range distance of $NN$ interactions.

Generally speaking, the pvCD-Bonn C potential can describe the phase shifts from Nijmegen PWA perfectly and generates the best properties of deuteron. Its $D$-state probability is higher than those from CD-Bonn potential and the latest N$^4$LO chiral potentials, whose $P_D$ are all around $4\%$. Actually, the $NN$ potentials with PV coupling usually generated a larger $P_D$, which was shown in Bonn A, B, C potentials and the work by Caia {\it{et al.}}~\cite{caia02}. 

\section{Summary and outlook}\label{sec.sum}
Based on the high-precision CD-Bonn potential, three revised $NN$ potentials were proposed, where the pion-nucleon coupling was taken as pseudovector form instead of the original pseudoscalar one, named as pvCD-Bonn A, B, C potentials. There were also $\omega,~\rho$, and two scalar mesons, $\sigma_1$ and $\sigma_2$, in these potentials besides pion. To describe the charge symmetry breaking and charge independence breaking of phase shifts from Nijmegen PWA more precisely, the coupling constants of $\sigma_1$ and $\sigma_2$ mesons were fitted independently at each partial wave. The strengths of one-pion-exchange component in  pvCD-Bonn A, B, C potentials were obviously distinguished in terms of different cut-off momenta.

The phase shifts from these three potentials at non-coupled channels were consistent with those from Nijmegen PWA. The only differences appear in the mixing parameters of coupled channels due to the different tensor components. The pvCD-Bonn C potential can describe all phase shifts from Nijmegen PWA very well in all channels up to $J=4$, which includes the strongest pion components. These potentials generate the similar properties of deuteron, such as $D/S$-state ratio, root-mean-square radius and the quadrupole moment, while the $D$-state probabilities from pvCD-Bonn A, B, C potentials were obviously different, which are $4.22\%,~5.45\%, ~6.05\%$, respectively. 

The original CD-Bonn potential has been applied to many aspects of nuclear physics and obtained great achievements. However it was very difficult to be used in relativistic framework due to the pseudoscalar pion-nucleon coupling, which generates a large attractive contribution from the nucleon-antinucleon excitation. Therefore, three charge-dependent one-boson-exchange potentials were proposed in this work with pion-nucleon pseudovector coupling by fitting the phase shifts from Nijmegen PWA. They could be widely adopted to calculate various nuclear many-body problems in the relativistic framework and to investigate the relativistic and tensor effects.  

	\section*{Acknowledgments}
		This work was supported in part by the National Natural Science Foundation of China (Grant No. 11775119, No. 11405116, and No. 11675083).

	\appendix
	 \section{The partial-wave-dependent parameters}\label{sec.app2}
		The coupling constants and masses of $\sigma_1$ and $\sigma_2$ mesons at various 
		partial waves up to $J=4$ for pvCD-Bonn A and pvCD-Bonn B potentials are listed in 
		Tables~ \ref{sigma-paramA} and  \ref{sigma-paramB}. The corresponding phase shifts 
		for $pp,~nn,$ and $np$ are shown in Tables~\ref{ppbonna}--\ref{npbonnb}.
		\begin{table}[tbp]
		  \centering
		    \caption{ Parameters of $\sigma_1$ and $\sigma_2$ adopted in pvCD-Bonn A potential. 
		    Blanks indicate the corresponding parameters taken as zero. Meson masses in unit 
		    MeV.}\label{sigma-paramA}
    		\setlength{\tabcolsep}{2mm}{
    		\begin{tabular}{c|cc|cc|cc|cc}
    		\toprule
           &$g^2_{\sigma_1}/4\pi$&$g^2_{\sigma_2}/4\pi$&$g^2_{\sigma_1}/4\pi$&$g^2_{\sigma_2}/4\pi$ 
           &$g^2_{\sigma_1}/4\pi$&$g^2_{\sigma_2}/4\pi$&$m_{\sigma_1}$&$m_{\sigma_2}$ \\
           \hline
		    $^1S_0$&   5.17     &  2.13    &  4.65    &   4.35   & 5.19     &  2.10    &  470  &  793  \\
		    $^3P_0$&  6.37     &  8.53    &  6.38    &   7.98   & 6.39     &  8.52    &  560  &  793  \\
		    $^1P_1$&            &          &  0.64    &   11.16  &          &          &  350  &  793  \\
		    $^3P_1$&  0.93      &  10.123  &  0.93    &   9.84   &  0.93    &  10.12   &  350  &  793  \\
		    $^3D_1$&            &          &          &          &          &          &       &       \\
		    $^3S_1$&            &          &  2.20    &   9.89   &          &          &  452  &  793  \\
		    $^1D_2$&  0.93      &  32.19   &   0.94   &  32.00   & 0.93     &  32.35   &  350  &  793  \\
		    $^3D_2$&            &          &   1.27   &  15.22   &          &          &  452  &  793  \\
		    $^3F_2$&  0.40      &  59.78   &   0.53   &  45.06   & 0.39     &  60.69   &  350  &  793  \\
		    $^3P_2$&  0.93      &  14.69   &   0.92   &  14.76   & 0.93     &  14.69   &  350  &  793  \\
		    $^1F_3$&            &          &   0.99   &          &          &          &  350  &       \\
		    $^3F_3$&  1.74      &  36.28   &   1.68   &  38.02   & 1.77     &  36.5    &  400  &  793  \\
		    $^3G_3$&            &          &   2.51   &          &          &          &  400  &       \\
		    $^3D_3$&            &          &   1.6    &  1.8     &          &          &  400  &  793  \\
		    $^1G_4$&  4.7       &          &   4.9    &          &  4.7     &          &  470  &       \\
		    $^3G_4$&            &          &   2.6    &          &          &          &  470  &       \\
		    $^3H_4$&  4.5       &          &   4.5    &          &  4.5     &          &  470  &       \\
		    $^3F_4$&  4.5       &          &   4.5    &          &  4.5     &          &  470  &       \\
		    \hline
		    &\multicolumn{2}{c|}{$pp$}&\multicolumn{2}{c|}{$np$}&\multicolumn{2}{c|}{$nn$}&
		    \multicolumn{2}{c}{}\\
		    \hline
		    \hline
		  \end{tabular}}
		\end{table}

\begin{table}[h]
	\centering
	\caption{ Parameters of $\sigma_1$ and $\sigma_2$ adopted in pvCD-Bonn B potential. Blanks indicate the corresponding parameters taken as zero. Meson masses in unit MeV.}\label{sigma-paramB}
	\begin{tabular}{c|cc|cc|cc|cc}
		\hline
		\hline
		&$g^2_{\sigma_1}/4\pi$&$g^2_{\sigma_2}/4\pi$&$g^2_{\sigma_1}/4\pi$&$g^2_{\sigma_2}/4\pi$ &$g^2_{\sigma_1}/4\pi$&$g^2_{\sigma_2}/4\pi$&$m_{\sigma_1}$&$m_{\sigma_2}$ \\
		\hline
		$^1S_0$&   5.19     &  5.30    &  4.89    &  11.54   & 5.20     &  5.58    &  470  &  1225 \\
		$^3P_0$&  5.07     &  43.42   &  5.06    &  34.10   &  5.09    & 42.87    &  520  &  1225 \\
		$^1P_1$&            &          &  0.74    &  89.73   &          &          &  350  &  1225 \\
		$^3P_1$&  2.36      &  52.46   &  2.42    &  40.79   &  2.38    &  52.09   &  424  &  1225 \\
		$^3D_1$&            &          &  1.88    &   1.30   &          &          &  452  &  793  \\
		$^3S_1$&            &          &  1.83    &   6.74   &          &          &  452  &  793  \\
		$^1D_2$&  2.19      &  208.07  &   2.20   &  206.78  & 2.23     &  32.35   &  400  & 1225  \\
		$^3D_2$&            &          &   1.45   &  21.33   &          &          &  424  & 1225  \\
		$^3F_2$&  1.87      &  23.97   &   1.86   &  24.09   & 1.86     &  23.79   &  424  &  793  \\
		$^3P_2$&  3.28      &  30.14   &   3.26   &  30.45   & 3.29     &  30.08   &  452  & 1225  \\
		$^1F_3$&            &          &   0.90   &          &          &          &  350  &       \\
		$^3F_3$&  3.01      &  40.82   &   2.96   &  41.32   &  3.02    &  40.90   &  452  &  793  \\
		$^3G_3$&            &          &   0.90   &          &          &          &  350  &       \\
		$^3D_3$&            &          &   0.80   &  5.54    &          &          &  350  &  793  \\
		$^1G_4$&  3.83      &          &   3.85   &          &  3.83    &          &  452  &       \\
		$^3G_4$&            &          &   3.60   &          &          &          &  470  &       \\
		$^3H_4$&  3.74      &          &   3.78   &          &  3.74    &          &  452  &       \\
		$^3F_4$&  3.74      &          &   3.78   &          &  3.74    &          &  452  &       \\
		\hline
		&\multicolumn{2}{c|}{$pp$}&\multicolumn{2}{c|}{$np$}&\multicolumn{2}{c|}{$nn$}&\multicolumn{2}{c}{}\\
		\hline
		\hline
	\end{tabular}
\end{table}

\begin{table}
  \centering
     \caption{$pp$ phase shifts in different partial waves, predicted by pvCD-Bonn A.}\label{ppbonna}
\setlength{\tabcolsep}{2.5mm}{
  \begin{tabular}{ccccccccccc}
     \hline
     \hline
   $T_\text{lab}$(MeV)&$^1S_0$&$^3P_0$&$^3P_1$&$^1D_2$&$^3P_2$&$\varepsilon_2$&$^3F_2$&$^3F_3$&$^1G_4$&$^3F_4$\\
       \hline
       1  & 32.82 & 0.14  & -0.08 &  0.00 &  0.02 &  0.00 &  0.00 & -0.00 &  0.00 & 0.00  \\
       5  & 54.84 & 1.60  & -0.90 &  0.04 &  0.24 & -0.05 &  0.00 & -0.01 &  0.00 & 0.00  \\    
      10  & 55.18 & 3.78  & -2.04 &  0.17 &  0.71 & -0.21 &  0.01 & -0.03 &  0.00 & 0.00  \\
      25  & 48.63 & 8.70  & -4.89 &  0.71 &  2.62 & -0.83 &  0.11 & -0.23 &  0.04 & 0.02  \\
      50  & 38.89 & 11.71 & -8.29 &  1.72 &  5.93 & -1.76 &  0.35 & -0.69 &  0.15 & 0.12  \\
      100 & 24.93 & 9.61  & -13.34&  3.74 & 10.89 & -2.74 &  0.82 & -1.49 &  0.42 & 0.50  \\
      150 & 14.78 & 4.73  & -17.53&  5.62 & 13.89 & -2.95 &  1.16 & -2.04 &  0.68 & 1.03  \\
      200 &  6.65 &-0.49  & -21.27&  7.25 & 15.67 & -2.74 &  1.36 & -2.43 &  0.94 & 1.63  \\
      250 & -0.20 &-5.60  & -24.68&  8.58 & 16.70 & -2.31 &  1.40 & -2.72 &  1.21 & 2.21  \\
      300 & -6.12 &-10.50 & -27.82&  9.59 & 17.20 & -1.77 &  1.29 & -2.96 &  1.48 & 2.74  \\
      \hline
     \hline
   \end{tabular}}

\end{table}

\begin{table}
  \centering
  \caption{$nn$ phase shifts in different partial waves, predicted by pvCD-Bonn A. }
\setlength{\tabcolsep}{2.5mm}{
  \begin{tabular}{ccccccccccc}
     \hline
     \hline
   $T_\text{lab}$(MeV)&$^1S_0$&$^3P_0$&$^3P_1$&$^1D_2$&$^3P_2$&$\varepsilon_2$&$^3F_2$&$^3F_3$&$^1G_4$&$^3F_4$\\
       \hline
       1  & 57.45 & 0.21  & -0.12 &  0.00 &  0.03 &  0.00 & 0.00 & -0.00 &  0.00 & 0.00   \\
       5  & 60.89 & 1.87  & -1.04 &  0.05 &  0.30 & -0.06 & 0.00 & -0.01 &  0.00 & 0.00   \\    
      10  & 57.71 & 4.14  & -2.24 &  0.19 &  0.82 & -0.22 & 0.02 & -0.04 &  0.00 & 0.00   \\
      25  & 49.01 & 9.00  & -5.12 &  0.75 &  2.84 & -0.86 & 0.11 & -0.24 &  0.04 & 0.02   \\
      50  & 38.59 & 11.71 & -8.54 &  1.78 &  6.23 & -1.79 & 0.35 & -0.71 &  0.16 & 0.12   \\
      100 & 24.38 & 9.43  & -13.61&  3.83 &  11.23& -2.75 & 0.82 & -1.50 &  0.42 & 0.51   \\
      150 & 14.15 & 4.51  & -17.82&  5.75 &  14.23& -2.93 & 1.17 & -2.05 &  0.69 & 1.06   \\
      200 & 6.00  &-0.76  & -21.57&  7.40 &  15.99& -2.70 & 1.36 & -2.43 &  0.95 & 1.66  \\
      250 & -0.84 &-5.87  & -24.99&  8.75 &  16.98& -2.25 & 1.40 & -2.71 &  1.22 & 2.25   \\
      300 & -6.78 &-10.70 & -28.14&  9.77 &  17.47& -1.70 & 1.29 & -2.95 &  1.49 & 2.78  \\
      \hline
     \hline
   \end{tabular}}
   
\end{table}

\begin{table}
  \centering
  \caption{$np~(T=1)$ phase shifts in different partial waves, predicted by pvCD-Bonn A. }
\setlength{\tabcolsep}{2.5mm}{
  \begin{tabular}{ccccccccccc}
     \hline
     \hline
   $T_\text{lab}$(MeV)&$^1S_0$&$^3P_0$&$^3P_1$&$^1D_2$&$^3P_2$&$\varepsilon_2$&$^3F_2$&$^3F_3$&$^1G_4$&$^3F_4$\\
       \hline
       1  & 62.07 & 0.18  & -0.11 &  0.00 &  0.02 &  0.00 & 0.00 & -0.00 &  0.00 & 0.00  \\
       5  & 63.67 & 1.63  & -0.93 &  0.04 &  0.28 & -0.05 & 0.00 & -0.00 &  0.00 & 0.00  \\    
      10  & 60.03 & 3.66  & -2.04 &  0.16 &  0.77 & -0.19 & 0.01 & -0.03 &  0.00 & 0.00  \\
      25  & 50.98 & 8.14  & -4.82 &  0.70 &  2.68 & -0.77 & 0.09 & -0.20 &  0.03 & 0.02  \\
      50  & 40.52 & 10.73 & -8.22 &  1.73 &  5.94 & -1.67 & 0.32 & -0.62 &  0.14 & 0.10  \\
      100 & 26.43 & 8.51  & -13.33&  3.83 & 10.82 & -2.64 & 0.77 & -1.38 &  0.39 & 0.47  \\
      150 & 16.34 & 3.66  & -17.59&  5.77 & 13.79 & -2.86 & 1.10 & -1.91 &  0.66 & 1.00  \\
      200 & 8.32  &-1.56  & -21.38&  7.44 & 15.56 & -2.65 & 1.27 & -2.27 &  0.94 & 1.59  \\
      250 & 1.58  &-6.64  & -24.84&  8.78 & 16.59 & -2.22 & 1.26 & -2.51 &  1.22 & 2.17  \\
      300 &-4.26  &-11.43 & -28.01&  9.79 & 17.11 & -1.68 & 1.09 & -2.68 &  1.50 & 2.69  \\
      \hline
     \hline
   \end{tabular}}
   
\end{table}
\begin{table}
  \centering
  \caption{$np~(T=0)$ phase shifts in different partial waves, predicted by pvCD-Bonn A. }
\setlength{\tabcolsep}{2.5mm}{
  \begin{tabular}{ccccccccccc}
     \hline
     \hline
   $T_\text{lab}$(MeV)&$^1P_1$&$^3S_1$&$\varepsilon_1$&$^3D_1$&$^3D_2$&$^1F_1$&$^3D_3$&$\varepsilon_3$&$^3F_3$&$^3G_4$\\
       \hline
       1  & -0.19 &147.69 & 0.10 & -0.01 &  0.01 & 0.00 &  0.00 &  0.00 & 0.00 & 0.00\\
       5  & -1.53 &118.00 & 0.59 & -0.18 &  0.23 &-0.01 &  0.00 &  0.01 & 0.00 & 0.00\\    
      10  & -3.13 &102.36 & 0.96 & -0.67 &  0.86 &-0.07 &  0.01 &  0.08 & 0.00 & 0.01\\
      25  & -6.51 & 80.23 & 1.24 & -2.77 &  3.76 &-0.42 &  0.06 &  0.56 &-0.05 & 0.17\\
      50  & -9.92 & 62.20 & 0.98 & -6.38 &  9.00 &-1.12 &  0.34 &  1.64 &-0.26 & 0.73\\
      100 &-14.49 & 42.50 & 0.17 & -12.11& 17.21 &-2.17 &  1.45 &  3.56 &-0.94 & 2.20\\
      150 &-18.23 & 29.99 &-0.55 & -16.22& 22.09 &-2.87 &  2.70 &  4.94 &-1.75 & 3.67\\
      200 &-21.56 & 20.63 &-1.19 & -19.22& 24.55 &-3.44 &  3.72 &  5.89 &-2.54 & 5.02\\
      250 &-24.60 & 13.08 &-1.76 & -21.43& 25.44 &-4.02 &  4.40 &  6.54 &-3.26 & 6.23\\
      300 &-27.40 &  6.73 &-2.29 & -23.07& 25.34 &-4.65 &  4.75 &  6.98 &-3.88 & 7.29\\
      \hline
     \hline
   \end{tabular}}   
\end{table}

\begin{table}
  \centering
     \caption{$pp$ phase shifts in different partial waves, predicted by pvCD-Bonn B. }
\setlength{\tabcolsep}{2.5mm}{
  \begin{tabular}{ccccccccccc}
     \hline
     \hline
   $T_\text{lab}$(MeV)&$^1S_0$&$^3P_0$&$^3P_1$&$^1D_2$&$^3P_2$&$\varepsilon_2$&$^3F_2$&$^3F_3$&$^1G_4$&$^3F_4$\\
       \hline
       1  & 32.80 & 0.14  & -0.08 &  0.00 &  0.01 &  0.00 &  0.00 & -0.00 &  0.00 & 0.00  \\
       5  & 54.85 & 1.59  & -0.90 &  0.04 &  0.22 & -0.05 &  0.00 & -0.01 &  0.00 & 0.00  \\    
      10  & 55.21 & 3.76  & -2.05 &  0.17 &  0.66 & -0.20 &  0.01 & -0.03 &  0.00 & 0.00  \\
      25  & 48.67 & 8.69  & -4.91 &  0.70 &  2.51 & -0.82 &  0.11 & -0.23 &  0.04 & 0.02  \\
      50  & 38.92 & 11.72 & -8.30 &  1.71 &  5.83 & -1.74 &  0.35 & -0.69 &  0.15 & 0.12  \\
      100 & 24.94 & 9.62  & -13.30&  3.74 & 10.95 & -2.75 &  0.83 & -1.50 &  0.41 & 0.50  \\
      150 & 14.77 & 4.73  & -17.49&  5.62 & 13.97 & -3.01 &  1.19 & -2.05 &  0.68 & 1.04  \\
      200 &  6.62 &-0.50  & -21.28&  7.23 & 15.69 & -2.86 &  1.39 & -2.40 &  0.94 & 1.63  \\
      250 & -0.25 &-5.60  & -24.77&  8.55 & 16.65 & -2.50 &  1.39 & -2.64 &  1.21 & 2.19  \\
      300 & -6.20 &-10.49 & -28.01&  9.60 & 17.15 & -2.03 &  1.22 & -2.81 &  1.47 & 2.68  \\
      \hline
      \hline
   \end{tabular}}
\end{table}

\begin{table}
  \centering
  \caption{$nn$ phase shifts in different partial waves, predicted by pvCD-Bonn B.}
\setlength{\tabcolsep}{2.5mm}{
  \begin{tabular}{ccccccccccc}
     \hline
     \hline
   $T_\text{lab}$(MeV)&$^1S_0$&$^3P_0$&$^3P_1$&$^1D_2$&$^3P_2$&$\varepsilon_2$&$^3F_2$&$^3F_3$&$^1G_4$&$^3F_4$\\
       \hline
       1  & 57.40 & 0.21  & -0.12 &  0.00 &  0.02 &  0.00 & 0.00 & -0.00 &  0.00 & 0.00   \\
       5  & 60.90 & 1.86  & -1.04 &  0.05 &  0.27 & -0.06 & 0.00 & -0.01 &  0.00 & 0.00   \\    
      10  & 57.75 & 4.12  & -2.24 &  0.18 &  0.76 & -0.22 & 0.01 & -0.04 &  0.00 & 0.00   \\
      25  & 49.06 & 8.98  & -5.13 &  0.74 &  2.72 & -0.85 & 0.11 & -0.24 &  0.04 & 0.02   \\
      50  & 38.66 & 11.72 & -8.53 &  1.77 &  6.14 & -1.78 & 0.35 & -0.71 &  0.16 & 0.12   \\
      100 & 24.44 & 9.44  & -13.53&  3.86 &  11.30& -2.76 & 0.84 & -1.52 &  0.42 & 0.52   \\
      150 & 14.21 & 4.51  & -17.73&  5.78 &  14.31& -2.99 & 1.20 & -2.06 &  0.69 & 1.06   \\
      200 & 6.05  &-0.77  & -21.52&  7.40 &  16.00& -2.82 & 1.39 & -2.42 &  0.95 & 1.65  \\
      250 & -0.80 &-5.88  & -25.02&  8.73 &  16.93& -2.44 & 1.39 & -2.65 &  1.22 & 2.22   \\
      300 & -6.75 &-10.69 & -28.26&  9.78 &  17.40& -1.96 & 1.21 & -2.81 &  1.49 & 2.71  \\
      \hline
     \hline
   \end{tabular}}
\end{table}

\begin{table}
  \centering
  \caption{$np~(T=1)$ phase shifts in different partial waves, predicted by pvCD-Bonn B.}
\setlength{\tabcolsep}{2.5mm}{
  \begin{tabular}{ccccccccccc}
     \hline
     \hline
   $T_\text{lab}$(MeV)&$^1S_0$&$^3P_0$&$^3P_1$&$^1D_2$&$^3P_2$&$\varepsilon_2$&$^3F_2$&$^3F_3$&$^1G_4$&$^3F_4$\\
       \hline
       1  & 62.10 & 0.18  & -0.11 &  0.00 &  0.02 &  0.00 & 0.00 & -0.00 &  0.00 & 0.00  \\
       5  & 63.68 & 1.62  & -0.93 &  0.04 &  0.25 & -0.05 & 0.00 & -0.00 &  0.00 & 0.00  \\    
      10  & 60.02 & 3.64  & -2.04 &  0.16 &  0.72 & -0.19 & 0.01 & -0.03 &  0.00 & 0.00  \\
      25  & 50.94 & 8.14  & -4.83 &  0.69 &  2.57 & -0.76 & 0.09 & -0.20 &  0.03 & 0.02  \\
      50  & 40.44 & 10.75 & -8.19 &  1.72 &  5.86 & -1.66 & 0.31 & -0.62 &  0.14 & 0.11  \\
      100 & 26.33 & 8.53  & -13.24&  3.83 & 10.89 & -2.66 & 0.77 & -1.38 &  0.39 & 0.48  \\
      150 & 16.26 & 3.67  & -17.51&  5.77 & 13.87 & -2.93 & 1.12 & -1.92 &  0.66 & 1.01  \\
      200 & 8.25  &-1.56  & -21.38&  7.42 & 15.59 & -2.80 & 1.30 & -2.27 &  0.94 & 1.59  \\
      250 & 1.53  &-6.64  & -24.95&  8.76 & 16.56 & -2.45 & 1.29 & -2.51 &  1.22 & 2.15  \\
      300 &-4.28  &-11.43 & -28.27&  9.82 & 17.07 & -1.99 & 1.10 & -2.68 &  1.50 & 2.65  \\
      \hline
     \hline
   \end{tabular}}
   
\end{table}
\begin{table}
  \centering
  \caption{$np~(T=0)$ phase shifts in different partial waves, predicted by pvCD-Bonn B.}\label{npbonnb}
\setlength{\tabcolsep}{2.5mm}{
  \begin{tabular}{ccccccccccc}
     \hline
     \hline
   $T_\text{lab}$(MeV)&$^1P_1$&$^3S_1$&$\varepsilon_1$&$^3D_1$&$^3D_2$&$^1F_1$&$^3D_3$&$\varepsilon_3$&$^3F_3$&$^3G_4$\\
       \hline
       1  & -0.19 &147.76 & 0.10 & -0.01 &  0.01 & 0.00 &  0.00 &  0.00 & 0.00 & 0.00\\
       5  & -1.50 &118.21 & 0.64 & -0.18 &  0.22 &-0.01 &  0.00 &  0.01 & 0.00 & 0.00\\    
      10  & -3.07 &102.65 & 1.08 & -0.68 &  0.85 &-0.07 &  0.01 &  0.08 & 0.00 & 0.01\\
      25  & -6.38 & 80.65 & 1.57 & -2.80 &  3.73 &-0.42 &  0.07 &  0.56 &-0.05 & 0.17\\
      50  & -9.75 & 62.72 & 1.65 & -6.43 &  8.97 &-1.11 &  0.38 &  1.62 &-0.26 & 0.72\\
      100 &-14.41 & 43.04 & 1.54 & -12.24& 17.22 &-2.13 &  1.48 &  3.51 &-0.93 & 2.18\\
      150 &-18.24 & 30.45 & 1.50 & -16.47& 22.12 &-2.82 &  2.69 &  4.86 &-1.74 & 3.65\\
      200 &-21.58 & 20.96 & 1.54 & -19.67& 24.58 &-3.40 &  3.69 &  5.79 &-2.56 & 5.02\\
      250 &-24.50 & 13.25 & 1.63 & -22.15& 25.48 &-3.99 &  4.40 &  6.42 &-3.45 & 6.27\\
      300 &-27.07 &  6.71 & 1.74 & -24.11& 25.38 &-4.65 &  4.80 &  6.85 &-4.07 & 7.38\\
      \hline
     \hline
   \end{tabular}}
\end{table}

\section{The formalism of one-boson-exchange potentials}\label{sec.app1}
The BbS equation in this work was solved in $LSJ$ basis. Therefore, the matrix elements of pvCD-Bonn potentials should be expressed in $LSJ$ basis. Actually, there are many classical literatures, which formulated the one-boson-exchange (OBE) potentials in $LSJ$ basis in detail such as Refs.~\cite{machleidt87,machleidt89,machleidt01}. From the quantum field theory, the OBE potentials are derived from the free nucleon scattering amplitudes as shown in Eq.~(\ref{NN-Amp}), where the Dirac spinor is obtained by solving a free Dirac equation and is normalized by $\bar{u}u=1$,
\begin{equation}
u(\mb{p},\lambda) = \sqrt{\frac{M+E}{2M}}
\left(\begin{array}{c}
1\\
\frac{\bm{\sigma}\cdot\mb{p}}{M+E} \\
\end{array}\right)|\lambda\rangle 
\end{equation}
where $|\lambda=\pm\frac{1}{2}\rangle$ is the wave function of spin. Here, the anti-nucleon freedom is neglected. The exact expressions of $\sigma$ meson, pion with pseudovector coupling, $\omega$ meson can be expanded within the Dirac spinor as  
\begin{align}
\nonumber
\bar{V}_\sigma (\mathbf{q}',\mathbf{q})&=-g_\sigma^2
\bar{u}(\mb{q}')u(\mb{q})\frac{1}{\mb{k}^2+m^2_\sigma}  
\bar{u}(-\mb{q}')u(-\mb{q}) \\
\label{NNS-Mom}
& = -g_\sigma^2\frac{W'W}{4M^2}
\left(1-\frac{\bm{\sigma}_1\cdot\mb{q}'}{W'}
\frac{\bm{\sigma}_1\cdot\mb{q}}{W}\right)
\left(1-\frac{\bm{\sigma}_2\cdot\mb{q}'}{W'}
\frac{\bm{\sigma}_2\cdot\mb{q}}{W}\right)
\frac{1}{\mb{k}^2+m_\sigma^2}, \\
\nonumber\\
\nonumber
\bar{V}^{\text{pv}}_{\pi}(\mathbf{q}',\mathbf{q})
&=-\frac{f^2_\pi}{m_\pi^2}\bar{u}(\mb{q}')\gamma^5
\bm{\gamma}\cdot \mb{k}u(\mb{q})\frac{1}{\mb{k}^2+m^2_\pi}  
\bar{u}(-\mb{q}')\gamma^5\bm{\gamma}\cdot \mb{k}u(-\mb{q}) \\
\nonumber
&=-\frac{f_\pi^2}{m_\pi^2}\frac{WW'}{4M^2}\left[
(E'-E+2M)\frac{\bm{\sigma}_1\cdot\mathbf{q}'}{W'}-
(E-E'+2M)\frac{\bm{\sigma}_1\cdot\mathbf{q}}{W}\right] \\
\label{NNP-pv-Mom}
&\quad \times\left[(E'-E+2M)\frac{\bm{\sigma}_2\cdot\mathbf{q}'}{W'}-
(E-E'+2M)\frac{\bm{\sigma}_2\cdot\mathbf{q}}{W}\right]
\frac{1}{\mb{k}^2+m_\pi^2}, \\
\nonumber  \\ \nonumber 
\bar{V}_\omega(\mb{q}',\mb{q}) &=g_\omega^2
\bar{u}(\mb{q}')\gamma^\mu u(\mb{q})\frac{1}{\mb{k}^2+m^2_\omega}  
\bar{u}(-\mb{q}')\gamma_\mu u(-\mb{q}) \\
\nonumber 
&=g_\omega^2\frac{W'W}{4M^2} \left[ 
\left(1+\frac{\bm{\sigma}_1\cdot\mb{q}'}{W'}
\frac{\bm{\sigma}_1\cdot\mb{q}}{W}\right) 
\left(1+\frac{\bm{\sigma}_2\cdot\mb{q}'}{W'}
\frac{\bm{\sigma}_2\cdot\mb{q}}{W}\right) \right. \\
\label{NNO-Mom}
&\quad +\left.\left(\frac{\bm{\sigma}_1\cdot\mb{q}'}{W'}
\bm{\sigma}_1+ \bm{\sigma}_1\frac{\bm{\sigma}_1 \cdot \mb{q}}{W}\right)
\left(\frac{\bm{\sigma}_2\cdot\mb{q}'}{W'}\bm{\sigma}_2+ \bm{\sigma}_2
\frac{\bm{\sigma}_2 \cdot \mb{q}}{W}\right)
\right]  \frac{1}{\mb{k}^2+m_\omega^2},
\end{align}
where, the wave functions of spin and isospin operator are omitted and $W' = E'+M$ and $W = E+M$. The expression of $\rho$ meson is more complicated than other mesons due to the tensor coupling part:
\begin{align}\nonumber
\bar{V}_\rho(\mb{q}',\mb{q}) =& \bar{u}(\mb{q}')
\left[g_\rho\gamma_\mu +\frac{f_\rho}{2M_p}(\gamma_\mu\bm{\gamma}\cdot\mb{k}
-k_\mu)\right] 
u(\mb{q})   \frac{1}{\mb{k}^2+m_\rho^2}       \\ \label{NNR-Mom}
&\quad~~\bar{u}(-\mb{q}')
\left[g_\rho\gamma^\mu +\frac{f_\rho}{2M_p}(\gamma^\mu\bm{\gamma}\cdot
\mb{k}-k^\mu)\right]u(-\mb{q}),
\end{align}
where $\gamma^\mu = (\gamma^0,\bm{\gamma})$ is the conventional gamma matrices and $k^\mu = (0,\mb{k})$ is the four-momenta between two interacting nucleons. The matrix elements of $\rho$ meson~\eqref{NNR-Mom} is divided into 3 pieces 
$\bar{V}_\rho = \bar{V}_{vv} + \bar{V}_{vt} + \bar{V}_{tt}$. The vector-vector coupling part is
\begin{equation*}
\bar{V}_{vv}(\mb{q}',\mb{q}) =g_\rho^2
\bar{u}(\mb{q}')\gamma^\mu u(\mb{q})\frac{1}{\mb{k}^2+m^2_\rho}  
\bar{u}(-\mb{q}')\gamma_\mu u(-\mb{q}),
\end{equation*}
which is identical to the $\omega$ meson potential~\eqref{NNO-Mom}. The vector-tensor coupling one is
\begin{align}
\nonumber 
\bar{V}_{vt}(\mb{q}',\mb{q})&=\frac{g_\rho f_\rho}{2M_p} 
[~\bar{u}(\mb{q}')\gamma_\mu u(\mb{q})\bar{u}(-\mb{q}')
(\gamma^\mu \bm{\gamma}\cdot \mb{k}-k^\mu) u(-\mb{q})  \\
\label{NNR-vt-Mom}
&+\bar{u}(\mb{q}')(\gamma_\mu \bm{\gamma}\cdot
\mb{k}-k_\mu)u(\mb{q})\bar{u}(-\mb{q}')\gamma^\mu u(-\mb{q})~] 
\frac{1}{\mb{k}^2+m_\rho^2}.
\end{align}
The tensor-tensor coupling one is written as,
\begin{align}
\label{NNR-tt-Mom}	
\bar{V}_{tt}(\mb{q}',\mb{q}) =\frac{f_\rho^2}{4M_p^2}
\bar{u}(\mb{q}')(\gamma_\mu \bm{\gamma}\cdot
\mb{k}-k_\mu)u(\mb{q})\frac{1}{\mb{k}^2+m_\rho^2}
\bar{u}(-\mb{q}')(\gamma^\mu \bm{\gamma}\cdot
\mb{k}-k^\mu) u(-\mb{q}).
\end{align}

The full pvCD-Bonn potentials~\eqref{NN-potential} are the sum of all meson contributions. In the above expressions, the spin structure from Dirac spinor are strongly dependent on the spin wave functions of in- and out- scattering states. The expectation values of spin structure are calculated within the spin wave function. To simplify the computational process, the helicity representation is adopted, where the spin is quantized along the direction of initial and final momenta, $|\lambda\rangle$.  In principle, there are 16 terms for the $\langle\lambda'_1\lambda'_2|V(\mathbf{q}', \mathbf{q})|\lambda_1\lambda_2\rangle$, since each helicity $\lambda$ can be $1/2$ or $-1/2$. Due to the parity conservation and time-reversal invariance of two identical fermions scattering,  only six matrix elements are independent:
\begin{equation}
\label{Hel-Mat}
\begin{aligned}
V^J_1(q',q) = \langle ++|V^J(q',q)|++\rangle, \quad 
V^J_2(q',q) = \langle ++|V^J(q',q)|--\rangle, \\
V^J_3(q',q) = \langle +-|V^J(q',q)|+-\rangle, \quad 
V^J_4(q',q) = \langle +-|V^J(q',q)|-+\rangle, \\
V^J_5(q',q) = \langle ++|V^J(q',q)|+-\rangle, \quad 
V^J_6(q',q) = \langle +-|V^J(q',q)|++\rangle, \\
\end{aligned}
\end{equation}
where the momentum angle dependence is integrated by
\begin{equation}
\label{Hel-Rot}
\langle\lambda'_1\lambda'_2|V^J(q',q)|\lambda_1\lambda_2\rangle 
=\int d\Omega d^J_{\lambda_1-\lambda_2,\lambda_1'-\lambda_2'}
\langle\lambda'_1\lambda'_2|V(\mathbf{q}', \mathbf{q})
|\lambda_1\lambda_2\rangle.
\end{equation}
The total angular momentum $J$ is conserved in two-nucleon scattering.  $d^J_{\lambda_1-\lambda_2,\lambda_1'-\lambda_2'}$ denote the reduced rotation matrices, which are expressed as,
\begin{equation}
\label{RotMat}
\begin{aligned}
d_{00}^J &= P_J, \\
(1+t)d_{11}^J&=\frac{P_{J-1}+JtP_J}{J+1}+P_J,  \\
(1-t)d_{-11}^J&=\frac{P_{J-1}+JtP_J}{J+1}-P_J,\\
\sin\theta d_{10}^J&=-\sin\theta d_{01}^J =\sqrt{\frac{J}{J+1}}(tP_J-P_{J-1}),
\end{aligned}
\end{equation}
where $t=\cos\theta = \hat{q}'\cdot\hat{q}$ and $P_J(t)$ are the Legendre polynomials. 

In center-of-mass frame, the helicity states for the nucleon with momenta $\mb{q}$ and $\mb{q}'$ can be constructed by  
\begin{equation}
\label{Hel-Stat-1}
\begin{aligned}
&\text{before scattering:} \quad \qquad
|+\rangle= \left(\begin{array}{c}
1 \\
0 \\\end{array}\right)
, \qquad \quad    |-\rangle =\left(\begin{array}{c}
0 \\
1 \\\end{array}\right), \\
&\text{after scattering:}\quad 
\langle+|=\left(\cos\frac{\theta}{2},~\sin\frac{\theta}{2}\right), \quad 
\langle-|=\left(-\sin\frac{\theta}{2},~\cos\frac{\theta}{2}\right),
\end{aligned}
\end{equation}
where $+$ represents $\lambda =+\frac{1}{2}$ and $-$ for $\lambda =-\frac{1}{2}$. The corresponding helicity states for momenta $-\mb{q}$ and  $-\mb{q}'$ are similarly shown as, 
\begin{equation}
\label{Hel-Stat-2}
\begin{aligned}
&\text{before scattering:} \quad \qquad |+\rangle= \left(\begin{array}{c}
0 \\
1 \\\end{array}\right)
, \qquad \quad    |-\rangle =\left(\begin{array}{c}
1 \\
0 \\\end{array}\right), \\
&\text{after scattering:}\quad 
\langle+|=\left(-\sin\frac{\theta}{2},~\cos\frac{\theta}{2}\right), \quad 
\langle-|=\left(\cos\frac{\theta}{2},~\sin\frac{\theta}{2}\right).
\end{aligned}
\end{equation}
With the expressions of reduced rotation matrices and helicity states, the integrals in Eq.~\eqref{Hel-Rot} contain the following seven types:
\begin{subequations}
	\label{Hel-Int}
	\begin{align}
	\label{Hel-Int0}
	I^{(0)}_J = & \int_{-1}^1 dt\frac{ P_J(t)}{q'^2+q^2-2q'qt+m_\sigma^2} 
	= \frac{Q_J(z)}{q'q}, \\
	\label{Hel-Int1}
	I^{(1)}_J = &\frac{1}{2q'q} \int_{-1}^1 dt\frac{ tP_J(t)}{z-t}~ \\
	\label{Hel-Int2}
	I_J^{(2)}=&\frac{1}{2q'q} \frac{1}{J+1}\int dt 
	\frac{JtP_J+P_{J-1}}{z-t},\\
	\label{Hel-Int3}	
	I^{(3)}_J=&\frac{1}{2q'q} \sqrt{\frac{J}{J+1}}\int_{-1}^1 dt
	\frac{tP_J-P_{J-1}}{z-t},\\
	\label{Hel-Int4}
	I^{(4)}_J=&\frac{1}{2q'q} \int_{-1}^1  dt \frac{t^2P_J}{z-t}, \\
	\label{Hel-Int5}	
	I^{(5)}_J=&\frac{1}{2q'q} \frac{1}{J+1}\int_{-1}^1  dt 
	\frac{Jt^2P_J+tP_{J-1}}{z-t},\\
	\label{Hel-Int6}
	I^{(6)}_J=&\frac{1}{2q'q} \sqrt{\frac{J}{J+1}}\int_{-1}^1  dt 
	\frac{t^2P_J-tP_{J-1}}{z-t}.
	\end{align}
\end{subequations}	
The first equation in Eq.~\eqref{Hel-Int0} can be worked out as the Legendre polynomials of the second kind 
\begin{equation}
Q_J(z) = \frac{1}{2}\int_{-1}^1 dt \frac{P_J(t)}{z-t},
\end{equation}
where $z = \frac{q'^2+q^2+m_a^2}{2q'q}$. The coupled channels in BbS equation are expressed more easily with the following linear combinations of matrix elements in Eq.~\eqref{Hel-Mat}. For spin-singlet ($S=0$) channel, there is 
\begin{equation}\label{Hel-Pot0}                   
^0V^J = V_1^J-V_2^J.
\end{equation}
The uncoupled spin-triplet ($S=1,~L=J$) channel is given as
\begin{equation}\label{Hel-Pot1} 
^1V^J = V_3^J-V_4^J ,
\end{equation}
while the coupled spin-triplet ($S=1$) channels have 
\begin{equation}
\label{Hel-Pot3} 
\begin{aligned}                      
^{12}V^J &= V_1^J+V^J_2, \\
^{34}V^J &= V_3^J+V_4^J,\\
^{55}V^J &= 2V_5^J, \\
^{66}V^J &= 2V_6^J. \\
\end{aligned}
\end{equation}

Therefore, the $NN$ potentials contributed from $\sigma$ meson in helicity basis are written as
\begin{equation}
\label{Pot-Sigma}
\begin{aligned}
^0V^J_\sigma(q',q)& = 
C_\sigma(F^{(0)}_\sigma I_J^{(0)}+F_\sigma^{(1)}I_J^{(1)}), \\
^1V_\sigma^J(q',q)&= 
C_\sigma(F^{(0)}_\sigma I_J^{(0)}+F_\sigma^{(1)}I_J^{(2)}), \\
^{12}V_\sigma^J(q',q) & = 
C_\sigma (F^{(1)}_\sigma I^{(0)}_J + F^{(0)}_\sigma I^{(1)}_J),\\
^{34}V_\sigma^J(q',q) &= 
C_\sigma (F_\sigma^{(0)}I^{(0)}_J+F^{(1)}_\sigma I_J^{(2)}), \\
^{55}V_\sigma^J(q',q) &= 
C_\sigma F_\sigma^{(2)}I_J^{(3)},\\
^{66}V_\sigma^J(q',q) & = ~^{55}V_\sigma^J(q,q').
\end{aligned}
\end{equation}
where the coefficient of $\sigma$ meson is
\begin{eqnarray}
	C_\sigma=\sqrt{\frac{M}{E'}}\sqrt{\frac{M}{E}}\frac{g_\sigma^2}{4\pi}\frac{1}{2\pi M^2}.
\end{eqnarray}
$~F^{(0)}_\sigma = -(M^2+E'E)~$,~$F^{(1)}_\sigma = q'q~$ and $~F_\sigma^{(2)} =M(E'+E)$.

The matrix elements of $NN$ interaction in helicity basis from pseudovector coupling of pion are
\begin{equation}
\label{Pot-Pion}
\begin{aligned}
^0V_\pi^J    &=~~C^{\text{pv}}_\pi(F_{\text{pv}\pi}^{(0)}I_J^{(0)}+
F_{\text{pv}\pi}^{(1)}I_J^{(1)})  \\
^1V_\pi^J    &=-C^{\text{pv}}_\pi(F_{\text{pv}\pi}^{(0)}I_J^{(0)} +
F_{\text{pv}\pi}^{(1)}I_J^{(2)})  \\
^{12}V_\pi^J &=~~C^{\text{pv}}_\pi(F_{\text{pv}\pi}^{(1)}I_J^{(0)}+
F_{\text{pv}\pi}^{(0)}I_J^{(1)})\\
^{34}V_\pi^J &=-C^{\text{pv}}_\pi(F_{\text{pv}\pi}^{(1)}I_J^{(0)} +
F_{\text{pv}\pi}^{(2)}I_J^{(2)})\\
^{55}V_\pi^J &=~~C^{\text{pv}}_\pi F_{\text{pv}\pi}^{(2)}I_J^{(3)}\\
^{66}V_\pi^J &=-C^{\text{pv}}_\pi F_{\text{pv}\pi}^{(2)} I_J^{(3)},
\end{aligned}
\end{equation}
with the coefficient
\begin{equation}
C^{\text{pv}}_\pi =\sqrt{\frac{M^2}{E'E}}
\left(\frac{f^2_\pi M^2}{\pi m_\pi^2}\right)
\frac{\vec{\tau}_1\cdot\vec{\tau}_2}{2\pi M^2}
\end{equation}
and
\begin{align*}
F_\pi^{(0)}&=E'E-M^2+(E'E+3M^2)\frac{(E'-E)^2}{4M^2}, \\
F_\pi^{(1)}&=-q'q+q'q\frac{(E'-E)^2}{4M^2}, \\
F_\pi^{(2)}&=-(E'+E)^2\frac{E'-E}{4M}.
\end{align*}
Similarly, the matrix elements from $\omega$ meson in helicity basis are
\begin{equation}
\label{Pot-Omega}
\begin{aligned}
	^0V_\omega^J    &=~~C_\omega(2E'E-M^2)I_J^{(0)},\\
	^1V_\omega^J    &=~~C_\omega(E'EI_J^{(0)}+q'qI_J^{(2)}), \\
	^{12}V_\omega^J &=~~C_\omega (2q'qI_J^{(0)}+M^2I_J^{(1)}), \\
	^{34}V_\omega^J &=~~C_\omega (q'qI_J^{(0)}+E'EI_J^{(2)}),\\
	^{55}V_\omega^J &=-C_\omega ME I_J^{(3)}, \\
	^{66}V_\omega^J &=-C_\omega ME'I_J^{(3)}.         
\end{aligned}
\end{equation}
with the coefficient
\begin{equation}
C_\omega=\frac{g^2_\omega}{4\pi^2M^2}
\sqrt{\frac{M^2}{EE'}}.
\end{equation}

There are two couplings between $\rho$ meson and nucleon, vector coupling and tensor coupling. Therefore, three components are generated in the matrix elements of $NN$ interaction from $\rho$ meson. The first one is obtained by the vector-vector coupling, which has the similar form as $\omega$ meson in Eq. (\ref{Pot-Omega}), except for the coefficient replaced by
\begin{equation}
C_{vv}=\frac{g^2_\rho}{4\pi}\frac{\vec{\tau}_1\cdot\vec{\tau}_2}{\pi M^2}
\sqrt{\frac{M^2}{EE'}}.
\end{equation}

The second component comes from vector-tensor coupling:
\begin{equation}
\label{Pot-Rho-vt}
\begin{aligned}
^0V^J_{vt}&=~~C_{vt}M[(q'^2+q^2)I^{(0)}-2q'qI_J^{(1)}], \\
^1V^J_{vt}    &=~~C_{vt}M(-(q'^2+q^2)I_J^{(0)}+2q'qI_J^{(2)}) ,  \\
^{12}V^J_{vt} &=~~C_{vt}M(6q'qI_J^{(0)}-3(q'^2+q^2)I_J^{(1)}),  \\
^{34}V^J_{vt} &=~~C_{vt}M(2q'qI_J^{(0)}-(q^2+q'^2)I_J^{(2)}),  \\
^{55}V^J_{vt} &=~~C_{vt} (E'q^2+3Eq'^2)I_J^{(3)},    \\
^{66}V^J_{vt} &=~~C_{vt} (Eq'^2+3E'q^2)I_J^{(3)},
\end{aligned}
\end{equation}
with
\begin{equation}
C_{vt}=\frac{g_\rho f_\rho}{4\pi M_p}
\frac{\vec{\tau}_1\cdot\vec{\tau}_2}{2\pi M^2}\sqrt{\frac{M^2}{EE'}}.
\end{equation}
where, $M_p$ is the proton mass as scaling mass in the tensor coupling between $\rho$ meson and nucleon.
The last contribution is denoted by tensor-tensor coupling:
\begin{equation}
\label{Pot-Rho-tt}
\begin{aligned}
^0V^J_{tt}=&C_{tt}\{(q'^2+q^2)(3E'E+M^2)I^{(0)}+[q'^2+q^2-2(3E'E+M^2)]q'q
I_J^{(1)}-2q'^2q^2I_J^{(4)}\}, \\
^1V^J_{tt} =&C_{tt}\{[4q'^2q^2+(q'^2+q^2)(E'E-M^2)]I_J^{(0)}+
2(E'E+M^2)q'qI_J^{(1)} \\
&-(q'^2+q^2+4E'E)q'qI_J^{(2)}-2q'^2q^2I_J^{(5)} \} ,  \\
^{12}V^J_{tt}=&C_{tt}\{[4M^2-3(q'^2+q^2)]q'qI_J^{(0)}+[6q'^2q^2-(q'^2+q^2)(E'E+3M^2)]I_J^{(1)}\\
&+2(E'E+M^2)q'qI_J^{(4)} \},  \\
^{34}V^J_{tt}=&C_{tt}\{-(q'^2+q^2+4E'E)q'qI_J^{(0)}-2q'^2q^2I_J^{(1)}
+[4q'^2q^2+(q'^2+q^2)(E'E-M)]I_J^{(2)} \\
&+2(E'E+M^2)q'qI^{(5)}_J \},\\
^{55}V^J_{tt}=&C_{tt}M\{ [E'(q'^2+q^2)+E(3q'^2-q^2)]I_J^{(3)}-2(E'+E)
q'qI^{(6)}_J\},      \\
^{66}V^J_{tt}=&C_{tt}M\{ [E'(q'^2+q^2)+E(3q^2-q'^2)]I_J^{(3)}-2(E+E')
q'qI^{(6)}_J\},
\end{aligned}
\end{equation}
with
\begin{equation}
C_{tt}=\frac{f_\rho^2}{4\pi M_p^2}
\frac{\vec{\tau}_1\cdot\vec{\tau}_2}{8\pi M^2}
\sqrt{\frac{M^2}{EE'}}.
\end{equation}

Furthermore, another representation, $|LSJ\rangle$, is more conveniently applied to the studies on $NN$ scattering and nuclear matter, where $L~,S$ denote the orbital angular momentum and spin, respectively. $J$ is the total angular momentum. In $|LSJ\rangle$ basis, the matrix elements are denoted as $V^{JS}_{L'L}=\langle L'SJ|V|LSJ\rangle$, which can be obtained by a unitary transformation from the ones in helicity basis given in~\eqref{Hel-Pot0}--\eqref{Hel-Pot3}. In spin-singlet ($S=0, ~L=J$) and uncoupled spin-triplet ($S=1,~L=J$) channels, the matrix elements in helicity basis and in $|LSJ\rangle$ are identical,
\begin{eqnarray}\label{JLS-Pot0}                   
V^{J0}_{JJ}&=&^0V^J,\\\nonumber
V^{J1}_{JJ}&=&^1V^J.  
\end{eqnarray}

The potentials with $|LSJ\rangle$ basis at coupled spin-triplet ($S=1$) channels are combined by
\begin{equation}
\label{JLS-Pot3} 
\begin{aligned}                      
V^{J1}_{J-1J-1} &= \frac{1}{2J+1}[J^{12}V^J+(J+1)^{34}V^J+\sqrt{J(J+1)} 
(^{55}V^J+~^{66}V^J)],   \\
V^{J1}_{J+1J+1} &= \frac{1}{2J+1}[(J+1)^{12}V^J+J^{34}V^J-\sqrt{J(J+1)} 
(^{55}V^J+~^{66}V^J)],  \\
V^{J1}_{J-1J+1} &= \frac{1}{2J+1}[\sqrt{J(J+1)}(^{12}V^J-~^{34}V^J)
-J^{55}V^J +(J+1)^{66}V^J)], \\
V^{J1}_{J+1J-1} &= \frac{1}{2J+1}[\sqrt{J(J+1)}(^{12}V^J- ~^{34}V^J)
+(J+1)^{55}V^J -J^{66}V^J)].  \\
\end{aligned}
\end{equation}


\begin{thebibliography}{99}
	\bibitem{Yukawa35} 	H. Yukawa, Proc. Phys. Math. Soc. Jpn. \textbf{17}, 48 (1935).
	
	\bibitem{bender03} M. Bender an P. -H. Heenen, Rev. Mod. Phys. {\bf 75}, 121 (2003).
	
	\bibitem{stone07} J. R. Stone, P. -G. Reinhard, Prog. Part. Nucl. Phys. {\bf 58}, 587 (2007).
	
	\bibitem{ring96} P. Ring, Prog. Part. Nucl. Phys. {\bf 37}, 193 (1996). 
	
	\bibitem{vretenar05} D. Vretenar, A. V. Afanasjev, G. A. Lalazissis, and P. Ring, Phys. Rep. {\bf 409}, 101 (2005).
	
	\bibitem{meng06} J. Meng, H. Toki, S. G. Zhou, S. Q. Zhang, W. H. Long, and L. S. Geng, Prog. Part. Nucl. Phys. {\bf 57}, 470 (2006).
	
	\bibitem{niksic11} T. Nik\v{s}i\'c, D. Vretenar, and P. Ring, Prog. Part. Nucl. Phys. {\bf 66}, 519 (2011).
	
	\bibitem{hamada62} T. Hamada and I. D. Johnston, Nucl. Phys. {\bf 34}, 382 (1962).
	
	\bibitem{reid68} R. V. Reid, Ann. Phys. (NY) {\bf 50}, 411 (1968).
	
	\bibitem{erkelenz74} K. Erkelenz, Phys. Rep. {\bf 13}, 191 (1974).
	
	\bibitem{machleidt87} R. Machleidt, K. Holinde, and Ch. Elster, Phys. Rep. {\bf 149}, 1 (1987).
	
	\bibitem{machleidt89} R. Machleidt, Adv. Nucl. Phys.  {\bf 19}, 189 (1989). 
	
	\bibitem{stoks94} V. G. J. Stoks, R. A. M. Klomp, C. P. F. Terheggen, and J. J. de Swart,  Phys. Rev. C {\bf 49}, 2950 (1994).
	
	\bibitem{wiringa95} R. B. Wiringa, V. G. J. Stoks, and R. Schiavilla, Phys. Rev. C {\bf 51}, 38 (1995).
	
	\bibitem{weinberg90} S. Weinberg, Phys. Lett. B {\bf 251}, 288 (1990).
	
	\bibitem{weinberg91} S. Weinberg, Nuclear Phys. B {\bf 363}, 3 (1991).
	
	\bibitem{weinberg92} S. Weinberg, Phys. Lett. B {\bf 295}, 114 (1992).
	
	\bibitem{ordonez94} C. Ord\'o\~nez, L. Ray, and U. van Kolck, Phys. Rev. Lett. {\bf 72}, 1982 (1994).
	
	\bibitem{ordonez96} C. Ord\'o\~nez, L. Ray, and U. van Kolck, Phys. Rev. C {\bf 53}, 2086 (1996).
	
	\bibitem{epelbaum98} E. Epelbaum, W. Gl\"{o}ckle, and U.-G. Mei\ss ner, Nucl. Phys. A {\bf 637}, 107 (1998).
	
	\bibitem{epelbaum00} E. Epelbaum, W. Gl\"{o}ckle, and U.-G. Mei\ss ner, Nucl. Phys. A {\bf 671}, 295 (2000).
	
	\bibitem{entem03} D. R. Entem and R. Machleidt, Phys. Rev. C {\bf 68}, 041001(R) (2003).
	
	\bibitem{epelbaum05} E. Epelbaum, W. Gl\"{o}ckle, and U.-G. Mei\ss ner, Nucl. Phys. A {\bf 747}, 362 (2005).
	
	\bibitem{entem15} D. R. Entem, N. Kaiser, R. Machleidt, and Y. Nosyk, Phys. Rev. C {\bf 91}, 014002 (2015).
	
	\bibitem{epelbaum15a} E. Epelbaum, H. Krebs, and U.-G. Mei\ss ner, Eur. Phys. J. A {\bf 51}, 53 (2015).
	
	\bibitem{epelbaum15b} E. Epelbaum, H. Krebs, and U.-G. Mei\ss ner, Phys. Rev. Lett. {\bf 115}, 122301 (2015).
	
	\bibitem{entem17} D. R. Entem, R. Machleidt, and Y. Nosyk, Phys. Rev. C {\bf 96}, 024004 (2017).
	
	\bibitem{reinert17} P. Reinert, H. Krebs, and E. Epelbaum, Eur. Phys. J. A {\bf 54}, 76 (2018).
	
	\bibitem{machleidt01} R. Machleidt, Phys. Rev. C {\bf 63}, 024001 (2001).
	
	\bibitem{caia02} G. Caia, J. W. Durso, Ch. Elster, J. Haidenbauer, A. Sibirtsev, and J. Speth, Phys. Rev. C {\bf 66}, 044006 (2002).
	
	\bibitem{lacombe75}  M. Lacombe {\it et al.}, Phys. Rev. D {\bf 12}, 1495 (1975).
	
	\bibitem{jackson75} A. D. Jackson, D. O. Riska, and B. Verwest, Nucl. Phys. A {\bf 249}, 397 (1975).
	
	\bibitem{drechsel92} S. K. D. Drechsel and L. Tiator, J. Phys. G {\bf 18}, 449 (1992).
	
	\bibitem{drechsel99} S. K. D. Drechsel, O. Hanstein, and L. Tiator, Nucl. Phys. A {\bf 645}, 145 (1999).
	
	\bibitem{fuchs98}C. Fuchs, T. Waindzoch, A. Faessler, and D. S. Kosov, Phys. Rev. C {\bf 58}, 2022 (1998).
	
	\bibitem{brockmann90} R. Brockmann and R. Machleidt, Phys. Rev. C {\bf 42}, 1965 (1990).
	
	\bibitem{blankenbecler66} R. Blankenbecler and R. Sugar, Phys. Rev. {\bf 142}, 1051 (1966).
	
	\bibitem{thompson70} R. H. Thompson, Phys. Rev. D {\bf 1}, 110 (1970).
	
	\bibitem{kadyshevsky68} V. G. Kadyshevsky, Nucl. Phys. B {\bf 6}, 125 (1968).
	
	\bibitem{haftel70} M. Haftel and F. Tabakin, Nucl. Phys. A \textbf{158}, 1 (1970).
	
	\bibitem{miller90} G. A. Miller, M. K. Netfkens and I. Slaus, Phys. Rep. \textbf{194}, 1 (1990).

	\bibitem{howell98} C. R. Howell, {\it et al.}, Phys. Lett. B \text{444}, 252 (1998).  

	\bibitem{gonzalez99} D. E. Gonz\'{a}lez Trotter {\it et al}., Phys. Rev. Lett. {\bf 83}, 3788 (1999).

	\bibitem{houk71} T. L. Houk, Phys. Rev. C {\bf 3}, 1886 (1971).

	\bibitem{rodning90} N. L. Rodning and L. D. Knutson, Phys. Rev. C {\bf 41}, 898 (1990).

	\bibitem{kermode83} M. W. Kermode, S. Klarsfeld, D. W. L. Sprung and J. P. McTavish, J. Phys. G \textbf{9}, 57 (1983).

	\bibitem{bishop79} D. M. Bishop and L. M. Cheung, Phys. Rev. A {\bf 20}, 381 (1979).

	\bibitem{martorell95} J. Martorell, D. W. L. Sprung, and D. C. Zheng, Phys. Rev. C {\bf 51}, 1127 (1995).

	\bibitem{gilman02} R. Gilman and F. Gross, J. Phys. G {\bf 28}, R37 (2002).
	
	\bibitem{bogner06} S. K. Bogner and R. J. Furnstahl, Phys. Lett. B \textbf{632}, 501 (2006).
	
	
	
	
	
\end{thebibliography}
\end{document}